\documentclass[aps,prx,twocolumn,nofootinbib,groupedaddress,superscriptaddress
,longbibliography]{revtex4-2}

\usepackage{amsmath,amssymb}
\usepackage{graphicx}
\usepackage{multirow}
\usepackage{bm}
\usepackage{mathtools}
\usepackage{dsfont}
\usepackage{amsfonts}
\usepackage{xcolor}
\usepackage[normalem]{ulem}
\usepackage{url}
\usepackage{wasysym}
\usepackage{bbm}
\usepackage{hyperref}
\usepackage{braket}
\usepackage{booktabs}

\usepackage[color=orange!60,textsize=scriptsize]{todonotes}
\newcommand{\equalcontrib}{These authors contributed equally.}

\newcommand{\ourtitle}{
Interaction-Driven Topological Transitions in Monolayer TaIrTe$_4$}

\allowdisplaybreaks

\begin{document}
\title{\textbf{\ourtitle}}

\author{Jiangxu Li}\thanks{\equalcontrib}

\affiliation{Department of Physics and Astronomy, The University of Tennessee, Knoxville, TN 37996, USA}

\author{Jian Tang}\thanks{\equalcontrib}
\affiliation{Department of Physics, Boston College, Chestnut Hill, MA, USA}

\author{Louis Primeau} 
\affiliation{Department of Physics and Astronomy, The University of Tennessee, Knoxville, TN 37996, USA}

\author{Thomas Siyuan Ding}
\affiliation{Department of Physics, Boston College, Chestnut Hill, MA, USA}

\author{Rahul Soni} 
\affiliation{Department of Physics and Astronomy, The University of Tennessee, Knoxville, TN 37996, USA}

\author{Tiema Qian}
\affiliation{Department of Physics and Astronomy and the California NanoSystems Institute, University of California, Los Angeles, 90095, CA, USA.}

\author{Kenji Watanabe}
\affiliation{Research Center for Electronic and Optical Materials, National Institute for Materials Science, 1-1 Namiki, Tsukuba 305-0044, Japan}

\author{Takashi Taniguchi}
\affiliation{Research Center for Materials Nanoarchitectonics, National Institute for Materials Science, 1-1 Namiki, Tsukuba 305-0044, Japan}

\author{Ni Ni}
\affiliation{Department of Physics and Astronomy and the California NanoSystems Institute, University of California, Los Angeles, 90095, CA, USA.}

\author{Adrian Del Maestro}
\affiliation{Department of Physics and Astronomy, The University of Tennessee, Knoxville, TN 37996, USA}
\affiliation{Min H. Kao Department of Electrical Engineering and Computer Science, The University of Tennessee, Knoxville, TN 37996, USA}

\author{Qiong Ma}
\affiliation{Department of Physics, Boston College, Chestnut Hill, MA, USA}
\affiliation{The Schiller Institute for Integrated Science and Society, Boston College, Chestnut Hill, MA, USA}

\author{Yang Zhang}
\affiliation{Department of Physics and Astronomy, The University of Tennessee, Knoxville, TN 37996, USA}
\affiliation{Min H. Kao Department of Electrical Engineering and Computer Science, The University of Tennessee, Knoxville, TN 37996, USA}

\begin{abstract}
Discovering materials that combine topological phenomena with correlated electron behavior is a central pursuit in quantum materials research. Monolayer TaIrTe$_4$ has recently emerged as a promising platform in this context, hosting robust quantum spin Hall insulator (QSHI) phases both within a single-particle gap and within a correlation-induced gap arising from van Hove singularities (vHSs), accessed via electrostatic doping. Its intrinsic monolayer nature offers exceptional tunability and the potential to realize a versatile array of interaction-driven topological phases. In this work, we combine theory and experiment to map the phase landscape of monolayer TaIrTe$_4$. Using Hartree-Fock calculations, we investigate 
the interaction-driven phase diagram near the vHSs under commensurate filling conditions. 
By systematically tuning the dielectric screening and strain, we uncover a rich set of ground states--including QSHI, trivial insulator, higher-order topological insulator, and metallic phase--among which are interaction-driven topological phase transitions. Experimentally, we perform both local and nonlocal transport measurements across a broad set of devices, which--due to unavoidable strain variations during fabrication-realize several phases consistent with theoretical predictions. Together, our results lay the groundwork for understanding correlation-driven topological phenomena in TaIrTe$_4$ and open new directions for engineering exotic quantum phases in low-dimensional materials beyond the limitations of moiré superlattices.

\end{abstract}

\maketitle

\let\oldaddcontentsline\addcontentsline
\renewcommand{\addcontentsline}[3]{}
\section{Introduction}
Enhanced electronic correlations, driven by high density of states (DOS) and suppressed kinetic energy, together with nontrivial band topology, provide an ideal platform for realizing exotic quantum phases.
Twisted and multilayer van der Waals (vdW) systems—such as graphene and transition metal dichalcogenides (TMDs)—have emerged as leading platforms in this pursuit, enabling the realization of topological flat bands through stacking and twist-angle engineering. This approach has proven remarkably successful, giving rise to a wide range of correlated phases, including experimentally observed (fractional) topological insulators~\cite{hsieh2008topological,chen2009experimental,xia2009observation,tang2024dual,fatemi2018electrically,kang2024evidence,kang2025time}, (fractional) Chern insulators~\cite{cai2023signatures,park2023observation,zeng2023thermodynamic,xu2023observation,lu2024fractional}, and, theoretically, the potential realization of non-Abelian states~\cite{reddy2024nonabelian,xu2025multiple,ahn2024non,wang2025higher}.

On the other hand, there remains strong motivation to discover natural materials in which topology and correlated electrons are intrinsically intertwined—potentially offering enhanced energy scales and eliminating the need for fine-tuned twist angles or complex fabrication procedures. This raises a key question: are there alternative platforms beyond moiré for realizing interaction-driven topological phases? One promising route is to induce correlations through van Hove singularities (vHSs) or flat bands arising from geometric frustration in topological materials, as exemplified by kagome~\cite{jiang2023flat,Farhang2023revealing,saykin2023high,farhang2023unconventional,teng2022discovery} and Lieb lattices~\cite{slot2017experimental,wei2019topological,cui2020realization}, as well as systems like bismuth~\cite{nayak2023visulizing,li2024spin} and FeGe~\cite{tan2025disordered}. Monolayer TaIrTe$_4$ emerges as a compelling candidate in this landscape: a two-dimensional topological insulator with vHSs located near its single-particle topological gap. 
Initially predicted to be a quantum spin Hall insulator (QSHI) based on its band structure~\cite{qian2014quantum,liu2025prediction,lai2024switchable}, recent experiments have not only confirmed this phase but also uncovered a correlated QSHI driven by electronic instabilities at the vHSs~\cite{tang2024dual}. Compared to natural three-dimensional bulk systems, intrinsic monolayer TaIrTe$_4$ offers key advantages: its chemical potential can be precisely tuned to the vHSs via electrostatic gating—crucial for accessing correlation-driven electronic phases. Furthermore, its exposed surface allows for dielectric environment engineering, while its monolayer nature endows it with the capacity to sustain large strains, providing additional tunability. These features establish monolayer TaIrTe$_4$ as a uniquely versatile platform for exploring the interplay between topology and strong correlations. Unraveling this interplay is essential for uncovering new quantum phases and enabling next-generation electronic applications.

In this work, we systematically investigate the topological and correlated phase diagram in monolayer TaIrTe$_4$ combining theoretical, numerical, and experimental approaches. Starting with density functional theory (DFT)~\cite{kohn1965self} and Lindhard susceptibility analysis, we identify a vHS-driven charge density wave (CDW) transition that leads to a 15$\times$1 superlattice structure. To further explore the interaction-driven phases, we construct a minimal eight-band tight-binding model and apply a Hartree-Fock mean-field approach~\cite{xu2024maximally,lu2023synergistic,huang2025doped}, focusing on the regime in which two additional electrons per supercell are introduced via electrostatic gating. Our calculations not only corroborate the previously reported correlated QSHI near vHSs~\cite{tang2024dual}, but also predict an extended phase diagram featuring multiple electronic states. These include correlated metallic phases, trivial insulators, and higher-order topological insulators (HOTI), arising from topological phase transitions controlled by dielectric environment and onsite interactions. 

To directly connect with experimentally accessible parameters, we further investigate uniaxial strain as an external tuning knob. Our theoretical results demonstrate that strain can induce topological phase transitions among the correlated states, offering a powerful route for experimentally engineering interaction-driven topological phases~\cite{tang2024dual,huang2025doped,zhao2020strain,liu2024controllable}. Complementing these theoretical insights, we fabricated and studied over 100 monolayer TaIrTe$_4$ devices, performing both local and nonlocal transport measurements to evaluate the correlated phases near vHSs. Strikingly, inherent strain variations introduced during device fabrication enable access to multiple distinct phases that are in agreement with our theoretical predictions.

Our study bridges theoretical modeling and experimental observation, offering deep insights into the complex interplay of topology and electron interactions in TaIrTe$_4$. These findings position monolayer TaIrTe$_4$ as a compelling alternative platform to moiré systems for exploring interaction-driven quantum phases.

\section{Single particle picture} \label{sec:SPP}
\subsection{Crystal Structure and Symmetries}\label{crystal}

\begin{figure}[htb!] 
    \centering
    \includegraphics[width=0.48\textwidth]{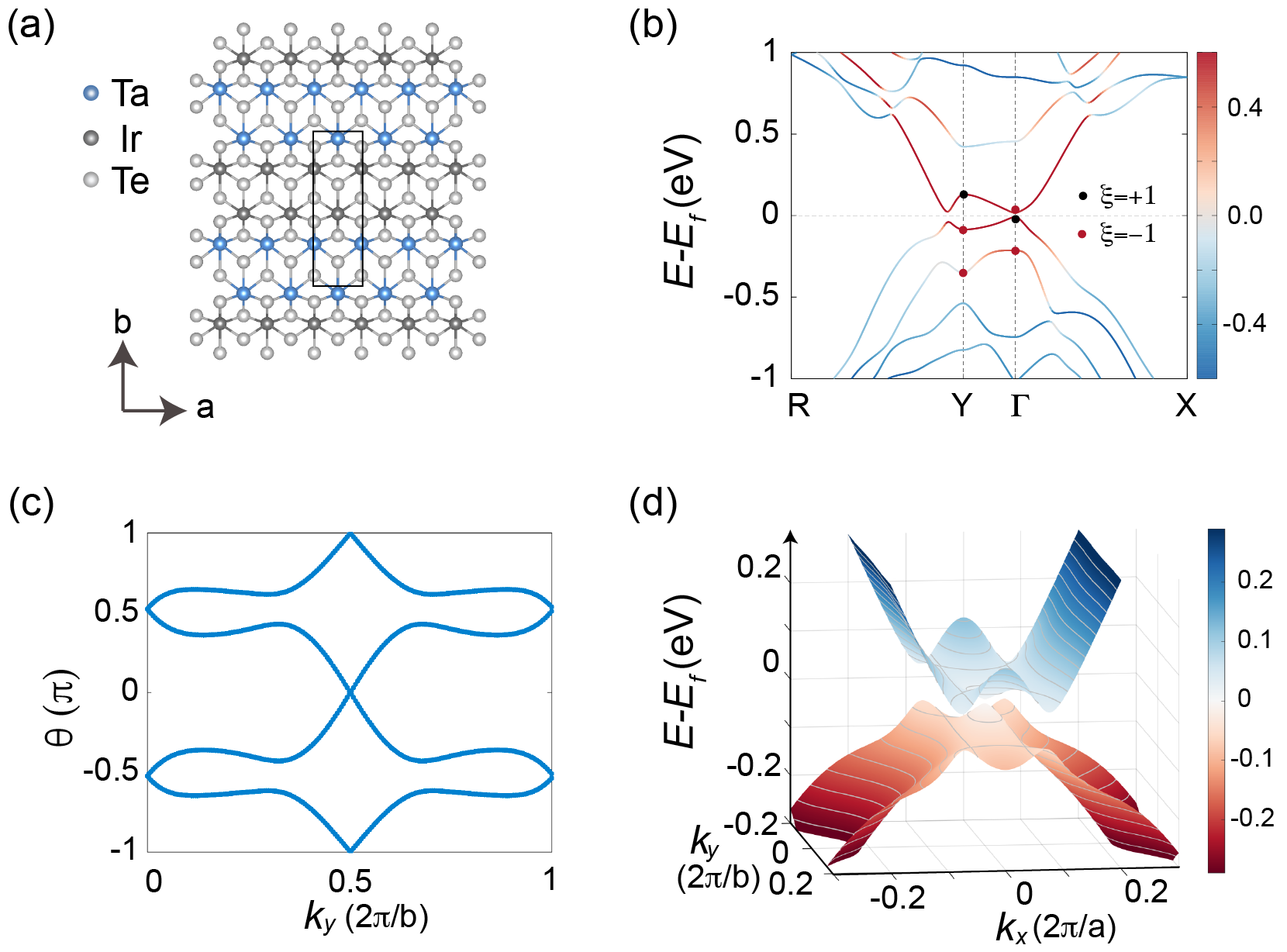}
    \caption{Band structures and topological properties of monolayer TaIrTe$_4$ by DFT calculation.
    (a) Schematic representation of the crystal structure, showing the arrangement of atomic sites.
    (b) DFT band structure along high-symmetry lines. The color gradient represents the weight differentiation between Ta and Ir+Te: red indicates a higher contribution from Ta, while blue signifies a dominant Ir and Te composition.
    The solid circles represent parities of bands 
    around the Fermi level.
    (c) Wilson loop of four valence bands near the Fermi level along the $k_y$ direction, indicating the topological nature of the bands.
    (d) 3D band structure of the lowest conduction and highest valence bands at $k_{xy}$ plane.} 
    \label{fig:fig0}
\end{figure}

As shown in Fig.~\ref{fig:fig0}(a), monolayer TaIrTe$_4$ has a space group of $P2_1/m$ ($SG$ 11) according to Ref.~\cite{liu2017van}.
Similar to monolayer 1T$^\prime$ WTe$_2$, 
Ta and Ir atoms are aligned along the $x$ direction, forming two one-dimensional metal chains.
The key symmetries in TaIrTe$_4$ include: 
lattice translations $t(\mathbf{R})$, with lattice vectors $\mathbf{R}$ = $m\mathbf{a}$+$n\mathbf{b}$, where $\mathbf{a}$ and $\mathbf{b}$ are the primitive lattice vectors, glide mirror symmetry $\tilde{\mathcal{M}}_x$ = $t(\mathbf{R}_y/2)\mathcal{M}_x$ with a regular reflection $\mathcal{M}_x$ normal to $\mathbf{a}$ and half-unit cell translation along $\mathbf{b}$, two-fold screw rotation $\tilde{C}_2$ = $t(\mathbf{R}_x/2)C_{2x}$ with a $180^\circ$ rotation and a fractional lattice translation along $\mathbf{a}$, time-reversal symmetry $\mathcal{T}$, and spatial inversion 
symmetry $\mathcal{P}$. The presence of Ta chains suggests a strong electronic anisotropy, and the Ta atoms dominate the Fermi level, as further confirmed by our DFT analysis, making a significant contribution to the system’s electronic susceptibility.

\subsection{DFT Band Structure and Topology} \label{sec:SPP-dft-bands}
First, based on the above structural information, 
we perform DFT calculation of the lattice and electronic structures of
2D TaIrTe$_4$ by Vienna \textit{ab} initio Simulation Package (VASP).
The relaxed lattice constants are $a = 3.816$\AA, $b = 12.55$\AA, and $c = 20$\AA~ (including vacuum), which agree well with experimental results~\cite{tang2024dual} and other calculations~\cite{liu2017van}.
Considering the effects of spin orbit coupling (SOC), we calculate the band structure along the high symmetry momenta with atom weights as shown in Fig.~\ref{fig:fig0}(b). The band gap is 21 meV at the $\Gamma$ point, in agreement with experimental measurements~\cite{tang2024dual}.

To analyze the atomic and orbital contributions to the bands around the Fermi level, we define a weight differentiation between Ta and Ir+Te atoms as $w=w_{Ta}-(w_{Ir}+w_{Te})$, marked
by color gradient, showing a higher contribution from Ta.
By analyzing the orbital distribution, we find that Ta-5$d$ orbitals dominate the orbital contributions around the Fermi level, specifically, the states of lowest conduction bands and highest valence bands are from the Ta's $d_{x^2-y^2}$,$d_{yz}$ and $d_{z^2}$ orbitals
 (see Supplementary Information Sec.\ref{Ssec:dft} for more details).
Hence, we choose two $d$ orbitals for our minimal tight-binding model in the Sec.~\ref{mini-tb}.
 
To determine the topological character of the bands, we calculate the parity eigenvalues ($\xi$) from the DFT wavefunctions at the time-reversal invariant momenta(TRIM), as indicated by the solid circles in Fig.~\ref{fig:fig0}(b).
The number of inversion-odd Kramers pairs
at the four TRIMs  $\Gamma$,
X, Y, and R are 22, 22, 21 and 22, respectively.
This leads to a QSHI with a $\mathbb{Z}_2$ invariant of $\nu_0=1$,
indicating that the band inversion occurs at Y.
As shown in Fig.~\ref{fig:fig0}(c), 
the Wannier center evolution of the four highest occupied valence bands
is plotted over the full Brillouin zone ($0\leq k_y\leq2\pi$).
Within the half Brillouin zone ($0\leq k_y\leq\pi$), the Wannier centers exhibit an odd number of crossings at each phase $\theta$, confirming the topological nature 
of the system with a $\mathbb{Z}_2$ invariant of 1.
The vHS can be clearly seen in Fig.~\ref{fig:fig0}(d).
Eight vHSs can be identified, with four in the conduction band and four in the valence band. These vHSs arise from the saddle points in the band structure, symmetrically distributed due to time-reversal symmetry $\mathcal{T}$ and glide mirror symmetry $\tilde{\mathcal{M}_x}$.

\begin{figure*}[htb!] 
    \centering
    \includegraphics[width=1.0\textwidth]{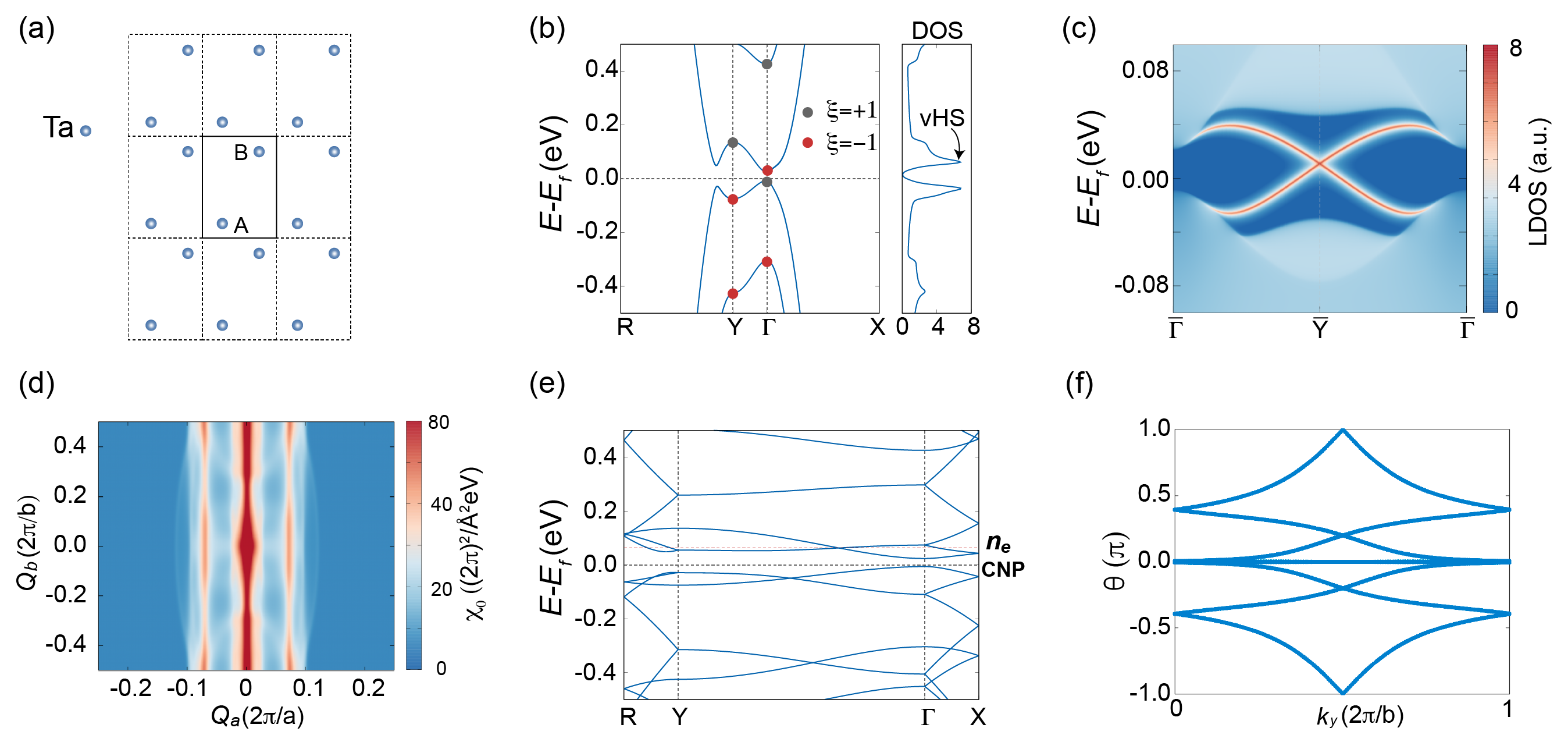}
    \caption{Band structures and topological properties of monolayer TaIrTe$_4$ by TB model.
    (a) Orbital positions.
    (b) Band structure along high-symmetry lines and the corresponding density of states.
    (c) Topological edge states along the [010] edge, showing the helical edge features.
    (d) Calculated charge susceptibility ($\chi_0$) of the conduction bands at the energy of vHS.
    Two peaks emerged at $Q_a = \pm 0.067$, connecting two vHS.
    (e) Folded band structures for a 15$\times$1 supercell configuration.
    (f) Wilson loop for the supercell, demonstrating the topological properties in the folded band structure.} 
    \label{fig:fig1}
\end{figure*}

\subsection{Tight-binding Band Structure and Topology}\label{mini-tb}

To enable subsequent many-body calculations in the CDW supercell, we construct a minimal tight-binding (TB) model guided by the DFT analysis above.
Since our focus lies on the low-energy physics near the vHS, we aim to develop a reduced model that faithfully captures the essential symmetries, topological features, and electronic structure of monolayer TaIrTe$_4$ near the Fermi level, while significantly reducing computational complexity.
Based on the symmetry analysis in Sec.~\ref{crystal}, we construct a tight-binding model with 8 bands for 4 orbitals with general Hamiltonian:

\begin{align}\label{eq:h0}
    \hat{H} = \sum_{\mathbf{R_1}, \mathbf{R_2}} \sum_{\alpha \beta s} t_{\alpha \beta}(\mathbf{R_1 - R_2}) a^{\dagger}_{\mathbf{R_1},\alpha s}a_{\mathbf{R_2},\beta s} 
\end{align}
where $\mathbf{R}_{1,2}$ are lattice vectors, $t$ is the hopping parameter including the chemical potential, 
$a_{\mathbf{R},\alpha s}^{\dagger}$($a_{\mathbf{R},\beta s}$) is the creation (annihilation) operator at lattice site $\mathbf{R}$ with orbital index $\alpha$ and spin index $s$.
Based on the crystal symmetry and DFT results reported above, we choose two $d$ orbitals for each sublattice $A$ and $B$ (see Fig.~\ref{fig:fig1}(a)), where the basis is ($a_{Ad_1s}$, $a_{Bd_1s}$,$a_{Ad_2s}$,
$a_{Bd_2s}$). In this basis, the matrix representations of the symmetry operations are 
$\Theta=is_y\tau_0\sigma_0\mathcal{K}$ for time reversal symmetry,
$P = s_0\tau_1\sigma_0$ for inversion symmetry,
and $M_x = is_x\tau_0\sigma_0$ for  nontranslational component of glide mirror symmetry $\tilde{\mathcal{M}}_x$. Therefore, the symmetry-allowed spin-orbital terms $H_{SOC}$ are $\lambda_{\pm}\sin(ak_x)s_z\tau_3(\sigma_0\pm\sigma_3)$.
We use particle swarm optimization to get the hopping parameters by fitting to the DFT bands and have tabulated them in Supplemental Material~\cite{supp}.

Figure~\ref{fig:fig1}(b) shows the band structure along
a high symmetry line and the DOS of the tight-binding model.
Due to the presence of both $\mathcal{P}$ and $\mathcal{T}$ symmetry,
each band is two-fold degenerate. Importantly, the model has a direct energy gap of
$\Delta E=23$ meV around the $\Gamma$ point and a vHS ($\mu_\textrm{vHS}$) 62 meV above the Fermi level,
both of which agree well with the above DFT results and experiment~\cite{tang2024dual}.
To confirm the topological properties,
we use the inversion operator $P$=$s_0\tau_1\sigma_0$ to get the
parities of all bands, which have been marked as solid circles at time-reversal invariant momenta in Fig.~\ref{fig:fig1}(b).
Comparing with the DFT results in Fig.~\ref{fig:fig0}(b), the parities of our model exhibit the same order around the Fermi level, indicating 
the same band inversion mechanism at the Y point.
The parity products of the occupied bands give us a $\mathbb{Z}_2$ invariant of $\nu_0=1$, as expected from DFT calculations.
More details on the band inversion mechanism and the parities of the DFT bands
can be found in Sec.~\ref{Ssec:model} of Supplemental Materials~\cite{supp}.
Fig.~\ref{fig:fig1}(c) shows the helical edge states along the high symmetry line of the [010] edge.
The helical edge states are degenerate at the $\overline{\textrm{Y}}$ point, connecting the conduction and valence bands and confirming
the band inversion at the Y point and the QSHI phase at the charge neutral point (CNP).

Furthermore, we calculate the charge susceptibility as shown in Fig.~\ref{fig:fig1}(d). It exhibits a pronounced peak at $Q_a=0.067\cdot 2\pi/a$, indicating that the electronic instability is driven by scattering between adjacent vHSs.
This suggests that the system favors a CDW state with a periodicity of approximately 15$\mathbf{a}$, directly linked to the nesting condition between vHSs. 

Therefore, the system can produce a 15$\times$1 CDW state by gating~\cite{tang2024dual}.
Fig.~\ref{fig:fig1}(e) shows the noninteracting folded band structure of a 15$\times$1 supercell.
Due to the band folding process, the dispersion of bands is much flatter around the Fermi level, especially at the vHS (marked with a red dashed line), where the electron density of states peaks and stronger electron interaction can be expected.
We also calculate the $\mathbb{Z}_2$ invariant and the Wannier center evolution along $\mathbf{k}_y$ of the supercell at CNP.
As shown in Fig.~\ref{fig:fig1}(f), the topology at the charge neutral point is the same as that of the unfolded bands, exhibiting a $\nu_0$=1 topological gap.

\begin{figure*}[htb!]
    \centering
    \includegraphics[width=1.0\textwidth]{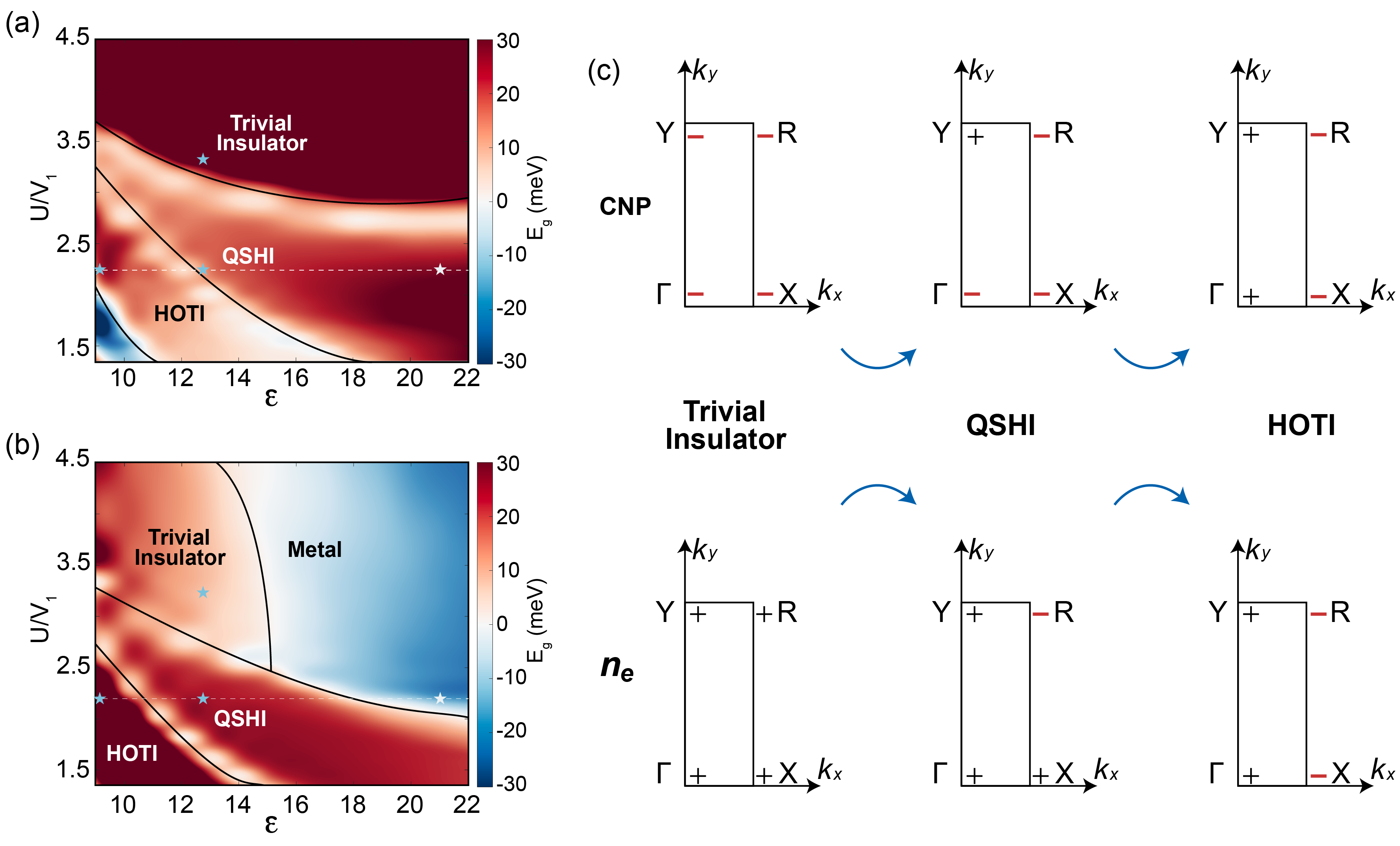}
    \caption{The Hartree-Fock mean-field phase diagram and associated band
    topology at filling of 0.13.
    (a) Phase diagram for CNP states, shown as a function of dielectric constant ($\epsilon$) and the relative interaction ratio ($U/V_1$).
    The color scale represents the energy gap between the 60th and 61st bands in meV, where red indicates an insulating state with a finite gap, and blue denotes a metallic state without a fully global gap. The white regions indicate phase boundaries. 
    (a) Phase diagram for the electron-doped states at $n = n_e$, with the gap defined between the 62nd and 63rd bands. White dashed line marks a reference value of $U/V_1$=2.25, while the solid black lines further outline the phase boundaries. Distinct phases are labeled: HOTI, QSHI, Trivial Insulator, and Metal.  
    (c) Parity products at the four time-reversal invariant momenta for the occupied bands: 1-60 for the CNP gap and 1-62 for the $n_e$ gap.
    The transitions between trivial insulator, QSHI, and HOTI phases are
    identified by changes in the parity products.}
    \label{fig:hfmf}
\end{figure*}

\section{Correlated electron states}\label{sec:hfmf}
\subsection{Hartree-Fock Mean-Field Calculations}

\begin{figure*}[htb!]
    \centering
    \includegraphics[width=0.9\textwidth]{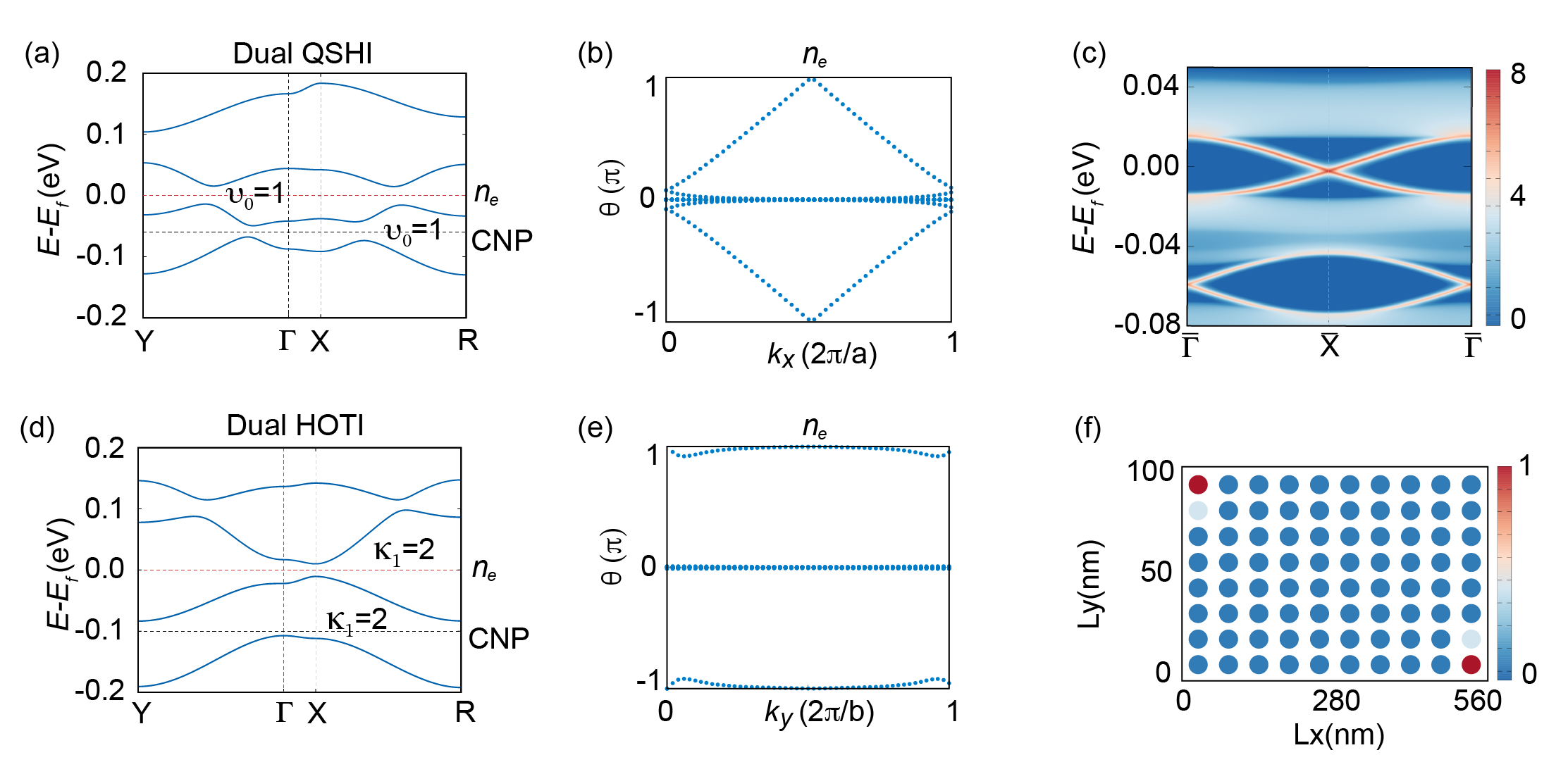}
    \caption{ Topological characterization of the dual QSHI and dual HOTI phases.
    (a-c) Band structure, Wilson loop evolution along $k_x$ for $n_e$ gap, [100] edge states for the dual QSHI phase, showing two $\mathbb{Z}_2$ topological 
    gaps at the $n_e$ and CNP levels.
    (d-f) The band structures along high-symmetry lines, evolution of the Wannier centers for $n_e$ gap, and spatial density profile of corner states for the HOTI phase, with both gaps exhibiting $\kappa_1$ = 2 higher-order topology.
    The color scale in (c) indicates the local density of states at the edge,
    while in (f) it represents the real-space density distribution of the corner
    states.}
    \label{fig:hfmf_p12}
\end{figure*}

Having established the framework to treat the vHS electronic instability on the single-particle level, we now consider how electron-electron correlations at the vHS will further modify the phases. 
To do that, we perform Hartree-Fock mean field calculations at the commensuate filling of 0.13 for the 15x1 supercell, that is two additional electrons per supercell from the CNP. 
We want to define this filling level as $n_e$ upfront to be connected with experimental observations later. 

We incorporate local Coulomb interactions using the extended Hubbard model, which includes both on-site Hubbard $U$ and density-density interaction $V$ terms:
\begin{align}
    \hat{H}_{\mathrm{int}}
        &=U\sum_{\mathbf{R}\alpha} \hat{n}_{\mathbf{R},\alpha\uparrow}\hat{n}_{\mathbf{R},\alpha\downarrow} \nonumber \\
        &+\sum_{(\mathbf{R_1}\alpha)\neq(\mathbf{R_2}\beta)}^{3rd NN}
        \sum_{ss^\prime}V_{\mathbf{R_1}\alpha,\mathbf{R_2}\beta} \hat{n}_{\mathbf{R_1,\alpha s}}\hat{n}_{\mathbf{R_2},\beta s'}~.
\end{align}
where subscript $\mathbf{R}$ labels a Bravais lattice site,  $\alpha$ and $\beta$ are composite sublattice and orbital indices, and $s$ and $s'$ are spin indices. $\hat{n}$ is the density operator, $U$ is the strength of the onsite Hubbard interaction, and $V$ is the amplitude of the density-density interaction between sites. A detailed derivation can be found in Sec.~\ref{Ssec:HFMF} of the Supplemental Materials~\cite{supp}.

Given the large number of 120 orbitals per supercell, the self-consistent mean-field equations require efficient numerical methods for convergence. We employ a matrix-product approach to compute the order parameters, leveraging the correlation function:
\begin{equation}
\langle (\mathbf{a}^\dagger)^T \mathbf{a}^T \rangle = \bar{U}^* \text{diag}(n_F(\lambda_i)) \bar{U}^T,
\end{equation}
where $\bar{U}$ represents the unitary matrix diagonalizing the mean-field Hamiltonian and $n_F(\lambda_i)$ is the Fermi-Dirac occupation function. This formulation enables rapid evaluation of self-consistent fields by exploiting batched eigenvalue decomposition, significantly accelerating the iterative procedure.
Additionally, a quasi-Newton acceleration method~\cite{Zhou2011aquasi} is employed, which allows for efficient convergence of the large Hartree-Fock fixed-point equations. The result is a self-consistent solution for the correlated phases of TaIrTe$_4$, which we analyze in the following section through a determination of the phase diagram and topological characterization.

\subsection{HFMF Phase Diagram}
To explore the role of electron interactions on both the CNP ($n=0$) and the doped 
states ($n=n_e$), we performed iterative self-consistent HFMF calculations
for the 15$\times$1 supercell at the filling of 0.13.
The converged phase diagrams, shown in Fig.~\ref{fig:hfmf}(a-b), are mapped as
a function of the dielectric constant $\epsilon$ and the interaction ratio $U/V_1$, where $U$ and $V_1 $ denote the onsite and nearest neighbor Coulomb repulsion, respectively. 
The formula relating $V_i$ and $\epsilon$ is given in the Supplementary Materials ~\cite{supp}.  
Although our calculations are carried out at a fixed filling of 0.13, corresponding to two extra electrons per supercell near the vHS,
the resulting self-consistent mean-field potential affects the entire spectrum.
As a result, both the CNP states and the $n_e$ states are renormalized by 
interactions and thus serve as fingerprints of the underlying many-body phase.
Fig.~\ref{fig:hfmf}(a) shows the phase diagram of the CNP gap, while 
Fig.~\ref{fig:hfmf}(b) shows the same for the $n_e$ gap.
These results demonstrate that electron correlations significantly modify both the intrinsic and emergent gaps.
Four distinct correlated electronic phases emerge in the phase diagram:
\begin{itemize}
\item \textbf{Quantum Spin Hall Insulator}: $\mathcal{PT}$ symmetric with nontrivial $\mathbb{Z}_2$ index $ \nu_0 = 1$, featuring edge states.
\item \textbf{Higher-Order Topological Insulator}: Identified by $\mathbb{Z}_4$ indicator $\kappa_1 = 2$, hosting corner states in real space.
\item \textbf{Trivial Insulator}: A trivial gapped state characterized by $\nu_0 = 0$ without band inversion, indicating the absence of topological edge or corner states.
\item \textbf{Metallic Phase}: A state where electron interactions fail to realize a fully gapped insulating phase, having nonzero DOS at the Fermi level.
\end{itemize}
As shown in Fig.~\ref{fig:hfmf}(c), different correlated phases can be analyzed by examining the parity products $\delta = \prod_{i=0}^n \xi_i$ of the $n$ occupied bands at the four time-reversal invariant momenta. 
As the relative interaction ratio varies, 
changes in the parity products indicate topological phase transitions.
The upper panel corresponds to the parity distribution for the CNP gap, while the lower panel represents that of the $n_e$ gap.

Specifically, when $U\gg V_1$, the onsite interaction takes the dominant role, pushing the system toward the atomic (trivial) insulator state with a large band gap. As $U/V_1$ decreases, the system transitions from a trivial insulator to a QSHI, with the band closing at the Y (R) point for CNP ($n_e$) gap, indicative of a parity change there.
Further decreasing the dielectric constant results in a transition from the QSHI to a HOTI, accompanied by a band gap closing and reopening at the $\Gamma$ (X) point for CNP ($n_e$) gap, signifying a parity change at  $\Gamma$ (X). Simultaneous band inversion at both R and X identifies the HOTI phase, characterized by the $\mathbb{Z}_4$ invariant $\kappa_1$=2.

These transitions demonstrate that electron correlation plays a fundamental role in modifying the topological properties of the system.
Furthermore, our Hartree-Fock calculations at 0.13 filling reveal that combined topological phases emerge depending on the topological nature of CNP and $n_e$ gaps as shown in Tab.~\ref{table:hfmf-phases}

\begin{table}[ht]
\centering
\caption{Classification of correlated phases based on the  CNP and $n_e$ states. The system belongs to the class AII of the Altland-Zirnbauer ten-fold symmetry classification. For insulating phases, Trivial insulator corresponds to $\mathbb{Z}_2$ invariant $\nu_0=0$, QSHI corresponds to $\nu_0=1$, HOTI corresponds to a $\mathbb{Z}_4$ invariant $\kappa_1=2$. Metal indicates no global gap.}
\begin{tabular}{p{4.2cm} p{2.2cm} c}
\hline
\textbf{Phase} & $n = 0$ & $n = $$n_e$ \\ \hline\hline
Dual QSHI & $\nu_0=1$  & $\nu_0=1$ \\[3pt]
Dual HOTI &  $\kappa_1=2$ & $\kappa_1=2$ \\[3pt]
Dual Trivial Insulator & $\nu_0=0$  &  $\nu_0=0$  \\[3pt]
HOTI + QSHI & $\kappa_1=2$  &  $\nu_0=1$ \\[3pt]
QSHI + Trivial Insulator& $\nu_0=1$ &  $\nu_0=0$ \\[3pt]
QSHI + Metal &  $\nu_0=1$ &  Metal \\[3pt]
Trivial Insulator + Metal &  $\nu_0=0$ &  Metal \\[3pt]
\hline\hline
\end{tabular}
\label{table:hfmf-phases}
\end{table}

To further characterize these phases, we analyze the associated band structures and topological nature at representative parameter points, as shown in Fig.~\ref{fig:hfmf}(a-b).

\begin{figure*}[htb!]
    \centering
    \includegraphics[width=0.9\textwidth]{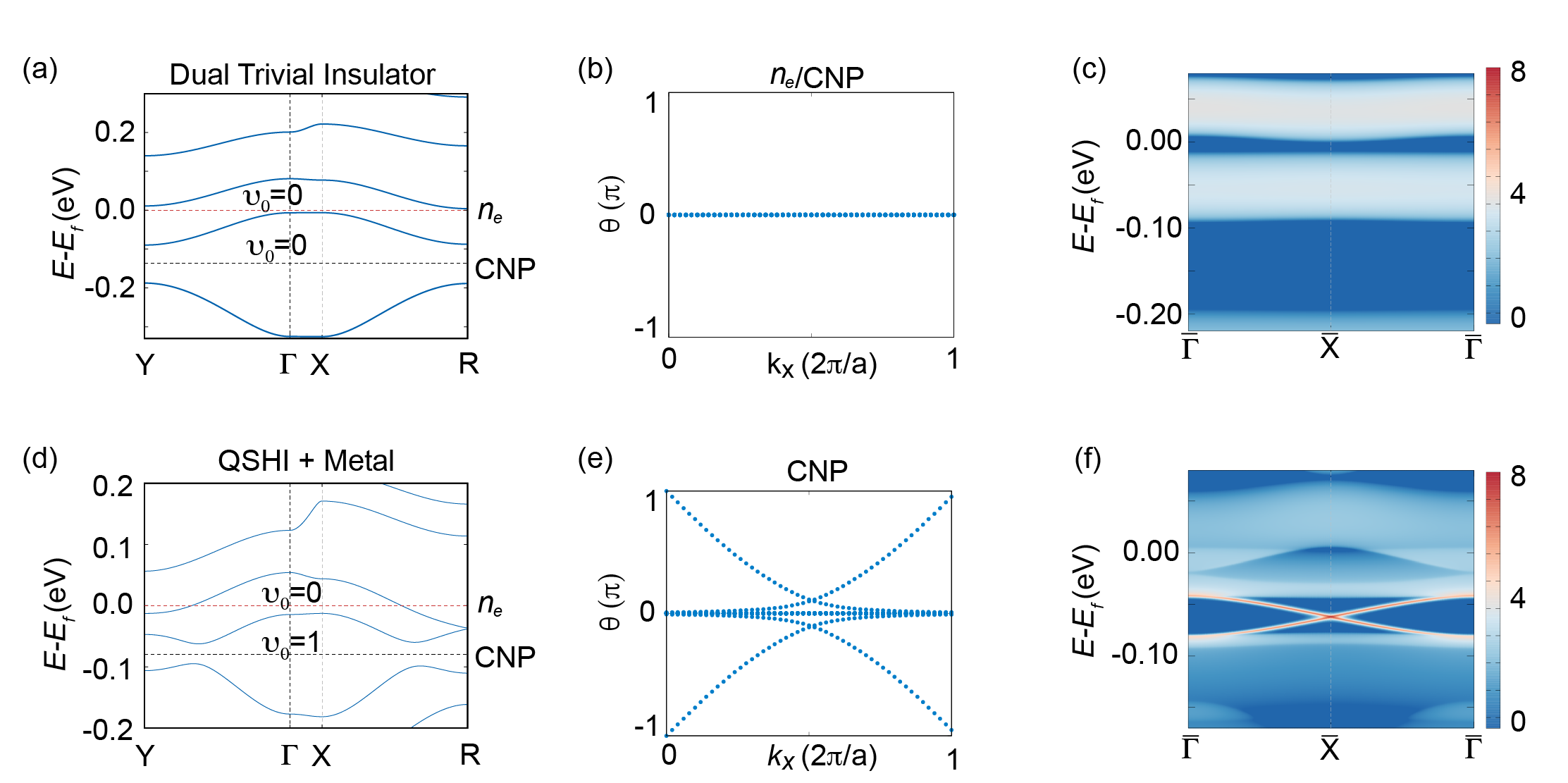}
    \caption{Topological band features of the dual trivial insulator 
    and QSHI+metal phases.
    (a-c) Band structure along high-symmetry lines, 
    Wilson loop evolution, and edge spectrum for the dual trivial insulator phase. Both the CNP and $n_e$ gaps are topological trivial with $\nu_0 = 0$, and no gapless edge states are observed.
    (d-f) Band structure, Wilson loop along $k_x$, and [100] edge states for the QSHI+metal phase. The CNP gap shows a nontrivial $\mathbb{Z}_2$ topology, 
    while the doping gap remains metallic without a global gap.
    }
    \label{fig:hfmf_p3}
\end{figure*}

\subsubsection{Dual QSHI}
To describe the dual QSHI phase, we take as a representative point $U/V_1 = 2.25$, $\epsilon = 13$. As shown in Fig.~\ref{fig:hfmf_p12}(a), the dual QSHI phase emerges when both the CNP gap and the $n_e$ gap exhibit a nontrivial $\mathbb{Z}_2$ topology.
The parity products at the TRIMs are shown in the middle panel of Fig.~\ref{fig:hfmf}(c).
Although the band inversion occurs at different momenta, 
both the CNP gap and the $n_e$ gap share the same $\mathbb{Z}_2$ topological
invariant $\nu$ = 1.
Fig.~\ref{fig:hfmf_p12}(b) shows the evolution of the Wannier centers of all occupied bands along the $k_x$ direction for the $n_e$ gap. The Wannier bands exhibit 
a $\pi$-phase difference at $k_x=\pi$, consistent with opposite parity at R and X. The Wannier bands along the $k_y$ direction exhibit the same behavior, as shown in Fig.~\ref{fig:dual_qshi}(c) in the Supplementary Materials~\cite{supp}, together indicating a clearly defined QSHI feature.
For the CNP gap, the Wannier band phases exhibit distinct phase jumps along different momentum directions, occurring at 
$k_x=0$ along $k_x$ and at $k_y=\pi$ along $k_y$ as illustrated in Fig.~\ref{fig:dual_qshi} of Sec.~\ref{Ssec:HFMF-results-dual-QSHI} in the Supplementary Materials~\cite{supp}.
As shown in Fig.~\ref{fig:hfmf_p12}(c), two distinctive helical edge states emerge in both the $n_e$ gap and CNP gap, confirming the topological properties of a dual QSHI. The gapless edge states are degenerate at $\bar{X}$ for the $n_e$ gap and $\bar{\Gamma}$ for the CNP gap, in agreement with the Wilson loop and parity product distributions. 
This phase is consistent with previously observed dual QSHI in experiments~\cite{tang2024dual}, further validating our theoretical approach.

\subsubsection{Dual HOTI}
As for the dual HOTI phase, we take the representative parameter point $U/V_1=2.25$, $\epsilon=9$ as an example.
Fig.~\ref{fig:hfmf_p12}(d) illustrates the dual HOTI phase, where both the CNP and $n_e$ gaps carry a higher-order topology characterized by the $\mathbb{Z}_4$ invariant $\kappa_1$ = 2.
As shown in Fig.~\ref{fig:hfmf}(c), the parity product distribution is the same for both gaps, where the band inversion occurs at both R
and X momenta.
Furthermore, the Wilson loop (Fig.~\ref{fig:hfmf_p12}(e)) along the $k_y$ direction shows a 
feature of a HOTI, with two bands at $\pm\pi$ emerging at the boundary, indicating that the average Wannier center is located at the boundary in real space.
Meanwhile, the Wilson loop phases along the $k_x$ direction remain zero, in agreement with the parity distributions [see more details in Sec.~\ref{Ssec:HFMF-results-dual-HOTI} of the Supplementary Materials~\cite{supp}].
To further confirm this phase, 
we construct a $10\times8$ rectangular superlattice with open boundary conditions and get the density distribution of the degenerate states in the $n_e$ gap as shown in Fig.~\ref{fig:hfmf_p12}(f). Two degenerate corner states are observed, obeying $C_{2z}$ symmetry. These corner states confirm the nontrivial topology associated with the HOTI phase, providing an additional layer of topological protection beyond conventional QSHI phases.

\subsubsection{Trivial Insulator and Metal}
Beyond the above exotic topological phases, Fig.~\ref{fig:hfmf_p3} illustrates the dual trivial insulator and metallic states. 
As shown in Fig.~\ref{fig:hfmf_p3}(a-c), the dual trivial insulator phase arises when both the CNP and $n_e$ gaps are topologically trivial ($\nu = 0$), leading two correlated insulating gaps without protected edge states. 
The parity products, shown in the left panel of Fig.~\ref{fig:hfmf}(c),
indicate the absence of band inversions for both CNP and $n_e$ gaps.
As the dielectric constant increases, the system transitions into a QSHI+metal phase (Fig.~\ref{fig:hfmf_p3}(d-f)), where the $n_e$ gap collapses. However, the CNP gap remains finite and retains its nontrivial topological character despite the metallic behavior in the doping region.
This demonstrates the crucial role of electron interactions in stabilizing correlated insulating phases, as the weakening of correlated effects leads to the emergence of metallic states [see more details in Sec.~\ref{Ssec:HFMF-results-metal-HOTI} of the Supplementary Materials~\cite{supp}].

These findings emphasize that the introduction of electron correlations not only stabilizes topological insulating states but also introduces new phases where different topological gaps coexist. Our results thus provide a theoretical framework to explain the experimentally observed dual QSHI and predict new correlated insulating and metallic phases in TaIrTe$_4$.

\subsection{Strain effects}

\begin{figure*}[htb!] 
    \centering
  \includegraphics[width=1.0\textwidth]{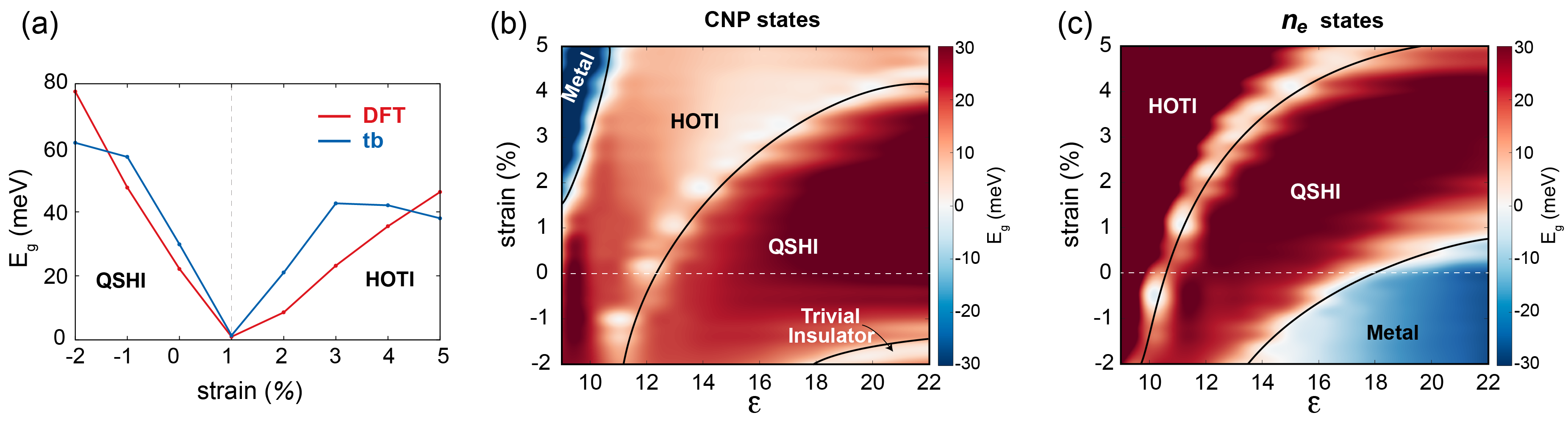}
    \caption{The phase transitions under strain.
    (a) The energy gap at different strains. The red lines represent
    the data from DFT calculations, while the blue lines are derived from the tight-binding model.
    (b) the HFMF phase diagram at filling $\nu$=0.13 as a function of the dielectric constant and the strain for the CNP states. Note that the dashed line in this figure matches the dashed line in Fig.~\ref{fig:hfmf}(a) and (b). 
    (c) the same diagram for the gaps between bands 62 and 63 for $n_e$ states. }
    \label{fig:fig4}
\end{figure*}

Thus far, our theoretical analysis has demonstrated a rich variety of correlated phases by tuning the interaction strength and dielectric screening. However, these parameters are not directly tunable in experiments. A more accessible experimental tuning parameter is uniaxial strain, which can influence both the single-particle electronic structure and the effective interaction in the system. Therefore, to systematically explore how strain modulates the correlated electronic phase diagram, we incorporate strain effects into our tight-binding model via the relation:
\begin{equation}
t_{ij}' = t_{ij} e^{-\beta (\delta / l)},
\end{equation}
where $t_{ij}'$ represents the modified hopping parameter under strain, $t_{ij}$ is the unstrained hopping amplitude, $\delta / l$ denotes the relative change in bond length due to strain, and $\beta$ is an empirical parameter that quantifies how sensitive the hopping amplitude is to lattice deformations.

To determine a suitable value for $\beta$, we apply uniaxial strain in our DFT calculations and extract the evolution of the energy gap at the CNP under strain. As shown in Fig.~\ref{fig:fig4}(a), when compressive strain is applied, the system remains a $\mathbb{Z}_2$ topological insulator, with an increasing band gap as the strain magnitude increases. In contrast, under tensile strain, a gap-closing and reopening transition occurs around 1\% strain, leading to the emergence of a HOTI phase (For more details on topological phase transitions by DFT see Sec.~\ref{Ssec:strain_effect} of the Supplementary Materials~\cite{supp}). 

By fitting the tight-binding model parameters to match the strain-dependent DFT gap closure, we obtain the blue curve in Fig.~\ref{fig:fig4}(a), with $\beta = 1.5$, which reproduces the topological transition in the DFT results.
Furthermore, to confirm the HOTI phase in our DFT calculations, we evaluate the parity products at the four time-reversal invariant momenta, namely $\Gamma$, X, Y, and R. The computed parity products at these points are ($-,+,-,+$), confirming a nontrivial $\mathbb{Z}_4$ invariant of $\kappa_1 = 2$. This result indicates that the system undergoes two band inversions, specifically at the $\Gamma$ and Y points, leading to the higher-order topology.

Using this strain-tuned tight-binding model, we compute the HFMF phase diagrams as a function of strain and dielectric constant, fixing $U/V_1 = 2.25$. The results are presented in Fig.~\ref{fig:fig4} (b-c), showing the strain-dependent phase evolution of the CNP gap and $n_e$ gap.

As shown in Fig.~\ref{fig:fig4}(b), the CNP gap exhibits multiple strain-induced phase transitions, including transitions between metal, HOTI, QSHI, and trivial insulator phases. Compared to the strain-induced effects on the $n_e$ states, strain has a broader impact on the CNP gap, significantly expanding the QSHI and HOTI phase regions.
The trivial insulator phase is characterized by parity products ($----$), while the QSHI phase has ($--+-$), indicating that band inversion occurs at the Y point. This observation is consistent with DFT results, confirming that electron interactions do not alter the intrinsic band inversion mechanism responsible for the formation of the QSHI at the CNP. 

For the $n_e$ gap, increasing tensile strain ($\delta > 0$) leads to a transition from a QSHI to a HOTI phase. The parity products at $\Gamma$, X, Y, and R change from ($+++-$) to ($+-+-$), indicating that the band inversion reoccurs at the X point. This transition causes the system's $\mathbb{Z}_2$ index $\nu_0$ to change from 1 to 0, while the $\mathbb{Z}_4$ invariant becomes $\kappa_1=2$, confirming the HOTI phase. Under compressive strain ($\delta < 0$), the $n_e$ gap increases for small $\epsilon$, reinforcing the QSHI phase (Fig.~\ref{fig:fig4}(b)). Additionally, for sufficiently large $\epsilon$, the QSHI phase transitions into a metallic state due to strong screening effects that suppress electron interactions. The interplay between strain, dielectric screening, and electron-electron interactions thus provides a powerful means of tuning topological phase transitions. Here, we also choose two representative phases to exhibit the 
strain effect on the correlated phase transitions.
For more detailed analysis of specific strain values see Sec.~\ref{Ssec:strain_effect_hfmf_1} and Sec.~\ref{Ssec:strain_effect_hfmf_2} of the Supplementary Materials~\cite{supp}. 

These results indicate that strain provides an effective tool for engineering correlated topological phases in TaIrTe$_4$, offering a pathway to control the emergence of QSHI, HOTI, and trivial insulating states through lattice deformation.

\section{Experimental assessment of phases}
\begin{figure*}[ht] 
    \centering
    \includegraphics[width=1.0\textwidth]{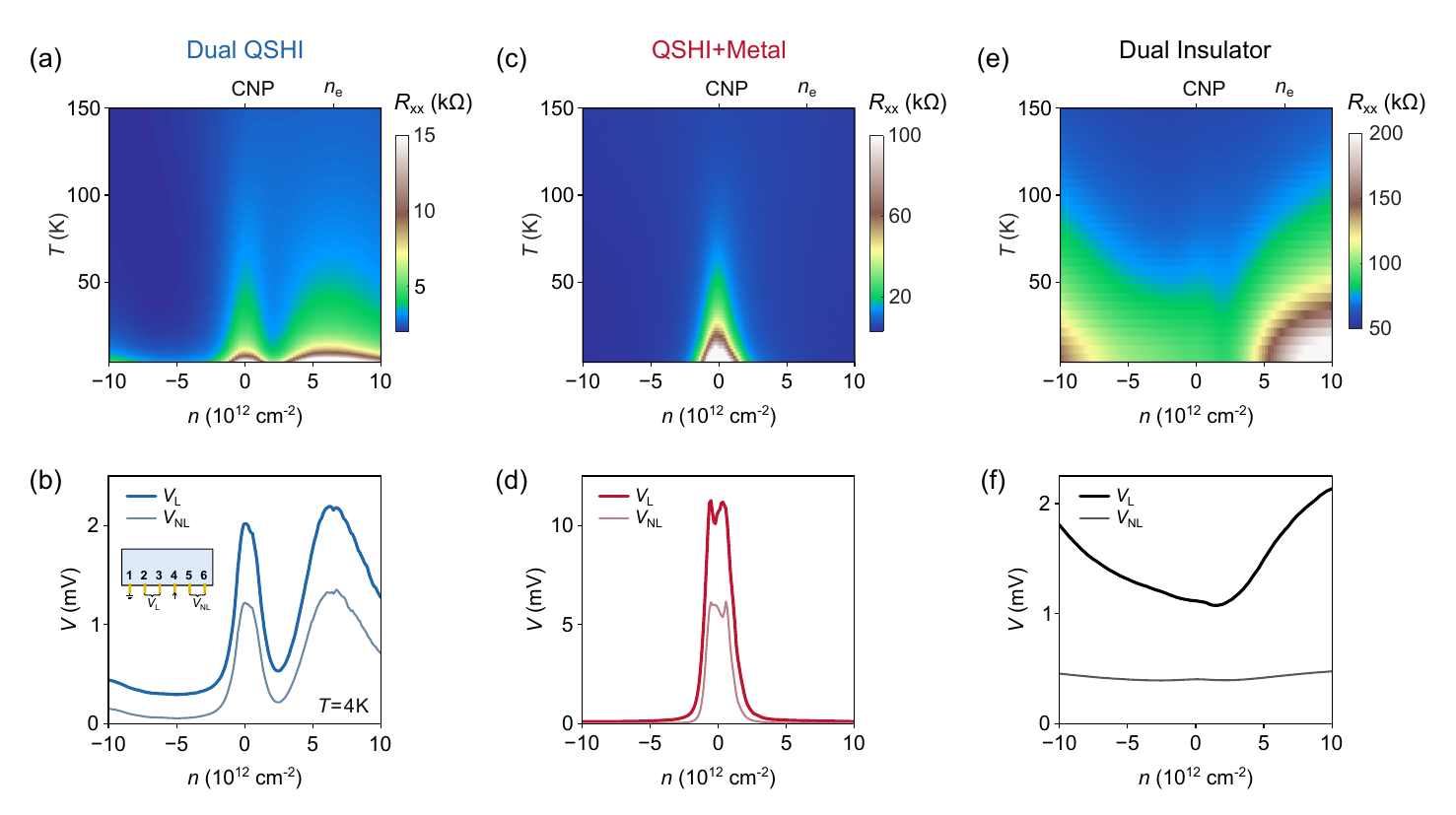}
    \caption{Three representative transport behaviors of monolayer TaIrTe$_4$ devices. 
    (a) Dual QSHI: Temperature ($T$)-dependent longitudinal resistance ($R_{xx}$) map as a function of carrier density ($n$), showing two pronounced resistive peaks at the CNP and at electron doping near $n_e = 6.5 \times 10^{12}$ cm$^{-2}$.
    (b) Local and nonlocal voltage responses versus $n$ at $T = 4$ K. Inset: schematic of the measurement setup, where current ($I_{xx}$) is applied between pins 4 and 1 (grounded), local voltage ($V_\mathrm{L}$) is measured between pins 3 and 2, and nonlocal voltage ($V_\mathrm{NL}$) between pins 5 and 6. Enhanced nonlocal signals are observed at both the CNP and the $n_e$ gap.
    (c) QSHI + metal: $R_{xx}$ map as a function of $n$ and $T$, showing a single resistive peak at the CNP. Resistance decreases upon electron or hole doping.
    (d) Corresponding local and nonlocal responses at $T = 4$ K, with strong nonlocal enhancement at the CNP gap.
    (e) Dual insulator: $R_{xx}$ map as a function of $n$ and $T$, displaying a weak resistive peak at the CNP. Resistance increases with both electron and hole doping. 
    (f) Corresponding local and nonlocal responses at $T = 4$ K, where the nonlocal response is suppressed across the entire doping range.}\label{fig:typical transport behaviors}
\end{figure*}

Uniaxial strain is, in principle, a controllable parameter in experiments, and monolayers can typically withstand substantial strain before mechanical failure. However, applying continuous and precisely controlled strain to small, fragile monolayer flakes remains experimentally challenging, requiring the careful selection of stretchable substrates and the development of compatible fabrication techniques. To date, only limited experimental work has demonstrated controllable and continuous strain tuning in thin layers~\cite{cenker2022reversible,cenker2023strain,hwangbo2024strain,liu2024continuously,cenker2025engineering}. While direct strain control represents a promising avenue for future studies, here we adopt a statistical approach to evaluate strain effects. During sample fabrications, TaIrTe$_4$ is picked up and transferred using flexible polymers (see Appendix~\ref{Sec:exp}) as part of the boron nitride encapsulation process. This procedure inevitably introduces unintentional, sample-dependent strain, which is difficult to control but expected to vary randomly across devices. By fabricating and characterizing a broad set of samples, we effectively probe a range of strain conditions. As we show below, this strategy enables access to multiple distinct regions of the phase diagram.

We fabricated dual-gated monolayer TaIrTe$_4$ devices and probed their electronic states using both local and nonlocal transport measurements, while continuously tuning the Fermi level from charge neutrality to the vicinity of vHSs. Although DFT calculations identify vHSs in both the conduction and valence bands, recent experiments~\cite{tang2024dual} indicate that only the conduction-band vHSs are fully accessible via electrostatic doping, leading to a correlated insulating gap at a doping density $n = n_e$. On the valence side, an onset of increasing resistance is observed at the highest accessible hole doping, suggesting that vHSs may also be present in the valence band but require stronger hole doping to reach. Therefore, our analysis focuses primarily on the charge neutrality point ($n = 0$) and the conduction-band vHS regime ($n = n_e$).

The experimental evaluation strategy proceeds as follows: (1) We first measure the temperature dependence of the longitudinal resistance to distinguish between insulating and metallic behaviors. (2) For samples exhibiting insulating behavior—indicative of a bulk gap—we perform nonlocal transport measurements to determine whether the nonlocal response is enhanced within the gap. Such enhancement suggests the presence of edge transport, consistent with QSHI regime. (3) Definitive confirmation of the QSHI requires the observation of a quantized longitudinal conductance of $2e^2/h$. However, as widely recognized in the community, achieving such quantization is experimentally challenging and requires ultra-short channel devices with nearly ideal contacts—conditions that are not practical for large-scale statistical studies. Therefore, we rely primarily on criteria (1) and (2) for statistical evaluation, while providing representative examples of quantized conductance from short-channel devices, with which the long-channel samples assigned to host QSHI states exhibit empirically consistent transport characteristics.

\textcolor{black}{From our previous DFT and HFMF calculations, we theoretically expect several possible correlated phases to emerge experimentally, including QSHI, HOTI, trivial insulator, and metallic phases. Experimentally, we can distinguish metallic from insulating phases directly through the temperature dependence of the longitudinal resistance $R_{xx} (T)$. Among insulating phases, QSHI states exhibit distinct quantized edge conductance, providing a clear transport signature. However, HOTI and trivial insulator phases both lack quantized edge transport, making it challenging to differentiate them using standard transport measurements alone. Thus, we collectively categorize these two indistinguishable states as \textit{insulators} in our experimental analysis.
Based on the above phase-classification strategy and by simultaneously considering the states at both the CNP and at $n_e$ doping level, we identify three distinct experimental phases, as summarized in Tab.~\ref{table:exp-phases}: (1) Dual QSHI, (2) QSHI + Metal, and (3) Dual Insulator. 
These experimentally observed states align closely with our theoretical predictions, and each regime is described in detail in the following sections.}




\begin{table}[ht]
\centering
\caption{Classification of correlated phases based on edge transport behaviors
at $n = 0$ for CNP phase and $n =$ $n_e$ for $n_e$ phase.}
\begin{tabular}{p{2.8cm} p{3cm} l}
\hline
\textbf{Phase} & $n = 0$ & $n =$ $n_e$ \\ \hline\hline
Dual QSHI & QSHI  & QSHI \\[3pt]
QSHI + Metal &  QSHI &  Metal \\[3pt]
\multirow{2}{*}{Dual Insulator} & HOTI or & HOTI or \\[3pt]
 &Trivial Insulator & Trivial Insulator \\[3pt]
\hline\hline
\end{tabular}
\label{table:exp-phases}
\end{table}


\subsection{Transport Characteristics of dual QSHI}

\begin{figure*}[htb!] 
    \centering
    \includegraphics[width=0.85\textwidth]{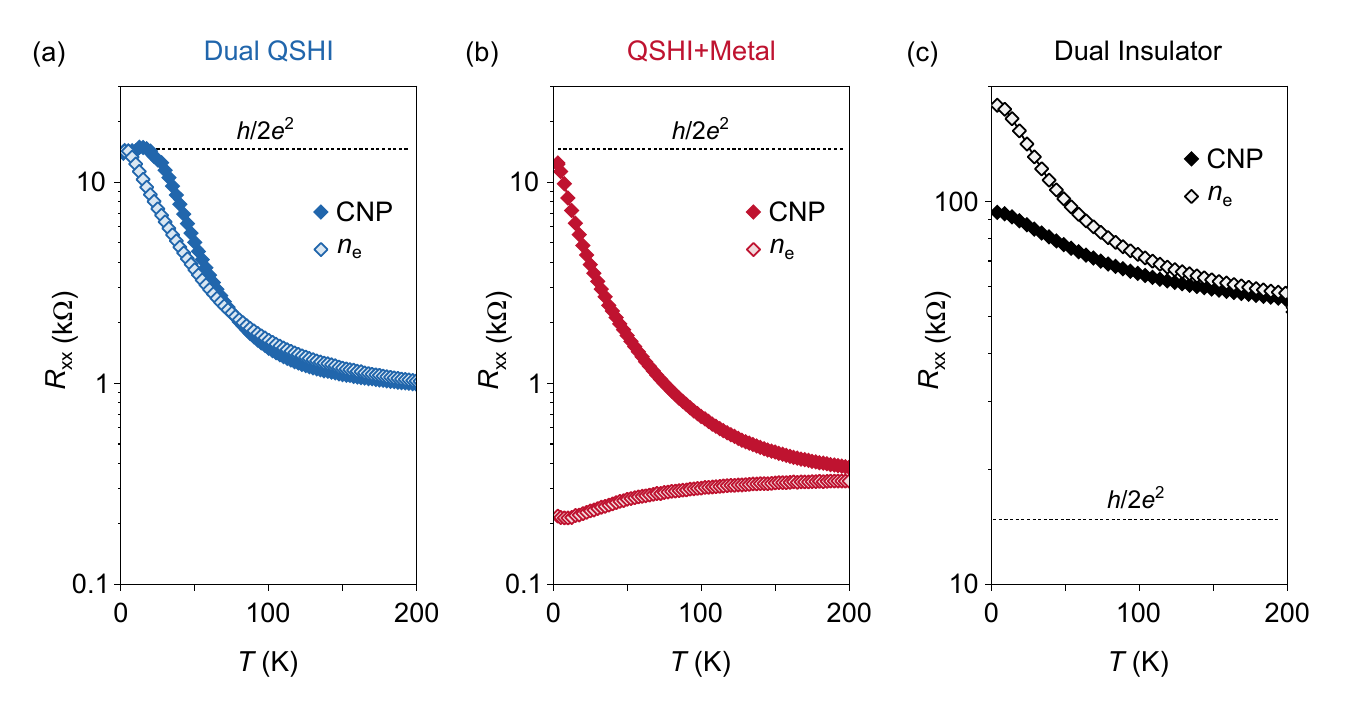}
    \caption{Temperature-dependent resistance of the CNP and $n_e$ phases in short-channel devices.
    (a) Dual QSHI: Both the CNP ($n = 0$) and the $n_e$ phase exhibit insulating behavior, with resistance saturating to the quantized value of $h/2e^2$ at low temperatures. Solid rhombus, CNP phase; hollow rhombus, $n = n_e$ phase.
    (b) QSHI + Metal: The CNP phase shows insulating behavior with resistance approaching $h/2e^2$, while the $n_e$ phase displays metallic behavior.
    (c) Dual Insulator: Both phases are insulating, but their resistances deviate significantly from the quantized value.
    }
    \label{fig:quantized}
\end{figure*}

We first describe the dual QSHI behavior, which is consistent with previous experimental reports~\cite{tang2024dual}. Among the 105 measured devices, approximately 35\% exhibit this qualitative behavior. The characteristic features of this type are as follows: at charge neutrality ($n = 0$), a resistance peak is observed, corresponding to the single-particle gap; at finite electron doping near $n_e \approx 6.5 \times 10^{12}$~cm$^{-2}$, a second resistance peak emerges, attributed to a vHS-induced electronic instability. These two resistance peaks are typically of comparable magnitude, on the order of 10 to 100~k$\Omega$, depending on the channel length. Temperature-dependent resistance maps reveal that both peaks correspond to insulating states, separated by a narrow metallic region, as illustrated in a representative device in Fig.~\ref{fig:typical transport behaviors}(a). In this device, thermal activation fitting yields an estimated charge-neutral gap of approximately 6.1~meV and an electron-doped gap of around 4.6~meV (see Sec.~\ref{Ssec:thermal_gap} of the Supplementary Materials~\cite{supp} for details).

In this type of device, we can further identify signatures of edge transport through nonlocal responses in a multi-terminal geometry. As shown in Fig.~\ref{fig:typical transport behaviors}(b), the system exhibits enhanced nonlocal signals—defined as the ratio between nonlocal and local voltage drops—within both gap regions. This behavior is consistent with dominant edge conduction, which confines current flow along the sample boundaries, resulting in enhanced nonlocal signals relative to local ones. To definitively confirm the topological nature of QSHI by the edge transport, short-channel devices (with channel lengths on the order of 100~nm) are required, as previously demonstrated for monolayer QSHI~\cite{tang2024dual,fatemi2018electrically}. This is because a short-channel device with a small size can effectively suppress bulk conduction arising from inhomogeneity and operate in a regime where the edge mean free path exceeds the channel length—that is, the edge ballistic transport regime. In one such device with a channel length of $L_{\mathrm{ch}} = 140$~nm, both resistance peaks saturate to the quantized value of $h/2e^2$ at low temperatures, as shown in Fig.~\ref{fig:quantized}(a). These results indicate that both the CNP and $n_e$ gaps correspond to QSHI.

\subsection{Transport Characteristics of QSHI + Metal}

Among the 105 measured devices, approximately 7\% exhibit the qualitative behavior of QSHI + metal, in which the CNP gap corresponds to a QSHI, while the $n_e$ state remains metallic rather than insulating. Results from a representative device displaying this behavior, which—unlike the previously discussed dual QSHI-exhibits a resistance peak only at the CNP, with no corresponding peak at $n = n_e$. Temperature-dependent measurements further confirm this distinction, revealing insulating behavior at the CNP and metallic behavior at $n_e$, as shown in Fig.~\ref{fig:typical transport behaviors}(c). For this device, thermal activation fitting yields a charge-neutral gap of approximately 28.0~meV, as estimated from the temperature-dependent resistivity at the CNP (Fig.~\ref{fig:thermal_gap}(a) in the Supplementary Materials~\cite{supp}), consistent with our DFT calculations (21 meV) in Sec~\ref{sec:SPP-dft-bands}. The QSHI nature of the CNP is further supported by nonlocal transport measurements (Fig.~\ref{fig:typical transport behaviors}(d)) and the observation of quantized conductance in a short-channel device (Fig.~\ref{fig:quantized}(b)), similar to the previously discussed case.

These experimental observations are in good agreement with the predictions from our HFMF phase diagram. The appearance of the QSHI + metal phase may be attributed to the screening-induced suppression of many-body electron-electron interactions at large dielectric constants. Under such conditions, the weakened correlations are insufficient to induce a global gap at this doping level, resulting in a metallic behavior consistent with a Fermi-liquid-like state. 
These results underscore the crucial role of electron interactions and dielectric screening in tuning the correlated topological phases in monolayer TaIrTe$_4$.

\subsection{Transport Characteristics of dual Insulator}

Interestingly, among the 105 measured devices, the majority (58\%) exhibit a dual insulator phase. In these monolayer TaIrTe$_4$ devices, the resistance peak at the CNP becomes less prominent compared to the steadily increasing resistance observed on both the electron- and hole-doped sides, as shown in Fig.~\ref{fig:typical transport behaviors}(e--f). Notably, no resistance peak is observed at $n = n_e$; instead, the resistance continues to rise throughout the accessible electron-doping range. Temperature-dependent measurements confirm that the entire doping range remains insulating [Fig.~\ref{fig:typical transport behaviors}(e)]. Moreover, no significant enhancement of nonlocal response is detected at either the CNP or $n = n_e$. In addition, in the short-channel limit, the resistance values [black curves in Fig.~\ref{fig:quantized}(c)] deviate significantly from the expected quantized edge conductance. Therefore, in this type of device, our observations suggest the absence of QSHI behavior at both $n = 0$ and $n = n_e$, despite the presence of insulating behavior at both doping levels.

Comparison with the theoretical phase diagram (Fig.~\ref{fig:hfmf} and Fig.~\ref{fig:fig4}) 
suggests two plausible scenarios for the experimentally observed dual insulator phase. Using the dual QSHI phase at $\epsilon = 13$ and $U/V_1 = 2.25$ as a reference point, 
we identify: (1) Increasing the interaction ratio to $U/V_1 \gtrsim 3.25$ at fixed $\epsilon = 13$ drives both the CNP and $n_e$ gaps into a trivial insulating phase (see Fig.~\ref{fig:hfmf}(b)).
(2) Alternatively, applying tensile strain beyond $\sim$3\% under moderate interaction strengths (still with $\epsilon = 13$, $U/V_1 = 2.25$) drives both gaps into the HOTI phase, as shown in Fig.~\ref{fig:fig4}. While HOTIs preserve nontrivial topology, their hallmark corner states are not detectable in edge transport measurements.
Since our measurements cannot distinguish between these two scenarios, we characterize this regime as a dual insulator phase, which lacks robust gapless edge modes both at $n = 0$ and $n =$ $n_e$.

Actually, the unavoidable strain in monolayer samples—arising from substrate interactions, device fabrication, or transfer processes—may naturally induce the transitions between the dual QSHI, QSHI+metal, and dual insulator phases observed in our experiments. In addition, sample-to-sample variations in dielectric layer thickness and gating environments can alter the effective screening strength, further shifting the system across interaction-driven phase boundaries. These extrinsic factors may account for the diversity of correlated phases realized across different devices.

\section{Conclusion and Discussion}

To summarize, this work integrates theoretical approaches with experimental measurements to investigate interaction-driven phase diagram in monolayer TaIrTe$_4$ that arise from the interplay between its QSHI topology and vHSs-induced correlations. Our theoretical framework includes DFT, tight-binding modeling, and Hartree-Fock mean-field calculations, which together reveal a rich landscape of phases—including HOTIs, trivial insulators, and metallic states—extending beyond the previously reported correlated QSHI near the vHSs. Notably, our theory predicts that when the Fermi level is tuned to the vHSs, electronic correlations can influence not only the correlated phase itself but also modify the topological character of the charge-neutral gap—a feature that is, in principle, not detectable through transport measurements, and calls for future spectroscopy investigation such as scanning tunneling microscopy.

On the experimental side, we investigate over 100 monolayer TaIrTe$_4$ devices using both local and nonlocal transport measurements. Unavoidable sample-to-sample strain variations and different dielectric screening introduced during fabrication, while unintentional, interestingly enable access to multiple interaction-driven phases. In addition to the correlated QSHI, we observe metallic and other insulating states near the vHSs. 
Each of these distinct regimes presents a promising avenue for future exploration. Our large-scale datasets enable a statistical classification of the observed behaviors—a level of experimental rigor and breadth that is unprecedented among monolayer systems, offering valuable guidance for ongoing and future investigations.

Overall, our findings establish a unified theoretical and experimental framework for correlated topological phenomena in TaIrTe$_4$ and related two-dimensional systems. The demonstrated tunability of topological states via electron interactions and strain offers exciting prospects for engineering quantum topological properties in van der Waals materials—beyond the constraints of moiré superlattices. 

Looking ahead, the interaction-induced strong superlattice potential provides a promising avenue for realizing exotic fractionalized phases—most notably, the elusive fractional topological insulator protected by time-reversal symmetry~\cite{levin2009fractional}, which has yet to be observed experimentally. Moreover, when the charge density wave instability associated with van Hove singularity is suppressed—via lattice strain, applied pressure, or enhanced Coulomb screening—superconductivity can emerge as a competing ground state~\cite{cai2019observation}. This continuous evolution from a correlated topological insulator to an intrinsic superconductor opens a novel path toward achieving topological superconductivity. Taken together, there are exciting opportunities for device architectures that harness the gate-tunable interplay of correlation, topology, and superconductivity—laying the foundation for integrated non-Abelian quantum platforms and superconducting quantum circuits~\cite{nayak2008non}.



\section{Acknowledgment}

We thank Cheng Xu and Ning Mao for their helpful discussions. J.L., L.P., A.D. and Y.Z. acknowledge support from the National Science Foundation Materials Research Science and Engineering Center program through the UT Knoxville Center for Advanced Materials and Manufacturing (DMR-2309083). The work done at Boston College was supported by the Air Force Office of Scientific Research (grants FA9550-22-1-0270 and FA9550-24-1-0117) and the Alfred P. Sloan Foundation. The single crystal growth and characterization of TaIrTe$_4$ at UCLA were supported by the U.S. Department of Energy (DOE), Office of Science, Office of Basic Energy Sciences under Award Number DE-SC0021117. K.W. and T.T. acknowledge support from the JSPS KAKENHI (Grant Numbers 21H05233 and 23H02052) , the CREST (JPMJCR24A5), JST and World Premier International Research Center Initiative (WPI), MEXT, Japan.



\bibliographystyle{apsrev}
\bibliography{ref}

\clearpage
\onecolumngrid
\begin{center}
\textbf{\large Supplemental Material for ``\ourtitle"}
\end{center}
\setcounter{section}{0}
\setcounter{figure}{0}
\setcounter{equation}{0}
\setcounter{table}{0}
\renewcommand{\thefigure}{S\arabic{figure}}
\renewcommand{\theequation}{S\arabic{equation}}
\renewcommand{\thesection}{S\arabic{section}}
\renewcommand{\thetable}{S\arabic{table}}

\appendix



\section{DFT calculations}\label{Ssec:dft}
\subsection{Methods}\label{Ssec:dft-methods}
To characterize the electronic structure and topological
properties of monolayer TaIrTe$_4$, we perform first-principles calculations
based on density functional theory (DFT) using the Vienna Ab initio Simulation Package (VASP) with the projector augmented wave (PAW) method.
Exchange-correlation effects are treated within the generalized gradient approximation (GGA) using the Perdew-Burke-Ernzerhof (PBE) functional.
A plane-wave cutoff energy of 500 eV is employed throughout the calculations.
The Brillouin zone (BZ) is sampled using a 16$\times$8$\times$1 Monkhorst-Pack k-point mesh, and a vacuum layer of 20 \AA~ is created along the $c$ direction to eliminate interlayer coupling. Structural relaxation is performed
with convergence criterion of $10^{-6}$ eV for the total energy and 
$10^{-3}$ eV/\AA ~for the ionic forces. The optimized in-plane lattice constants
are $a=3.816$\AA ~and $b=12.55$ \AA,~agreeing well with previous experimental
and calculations.

\subsection{Band Structure}\label{Ssec:dft-bands}
\begin{figure*}[htb!] 
    \centering
    \includegraphics[width=0.7\textwidth]{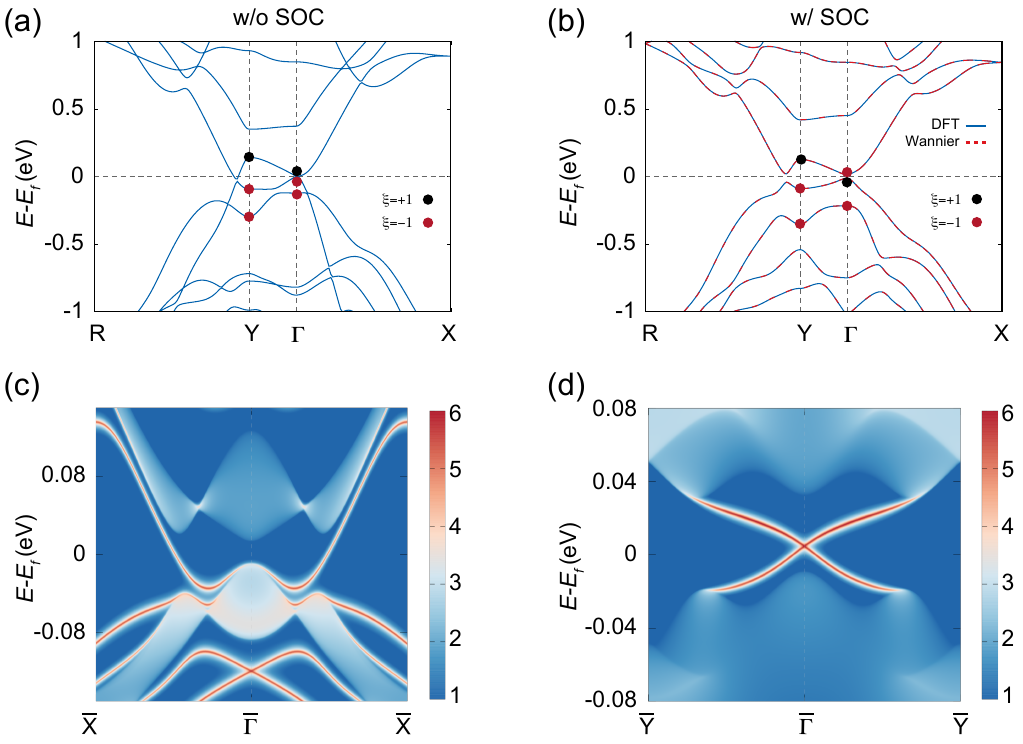}
    \caption{DFT band structures and topological edge states of monolayer TaIrTe$_4$.
    (a) DFT band structure without SOC. Parity eigenvalues ($\xi=\pm 1$) at the TRIM
    points $\Gamma$ and Y indicate idential band inversion pattern ($-,-,+$), resulting
    in a trivial $\mathbb{Z}_2$ topological phase.
    (b) DFT band structure with SOC. The parity eigenvalues at $\Gamma$ change to 
    $(-,+,-)$, resulting in a single band inversion and a nontrivial $\mathbb{Z}_2$ invariant $\nu_0=1$. The Wannier-interpolated bands (red dashed lines) closely reproduce the DFT results (blue lines), confirming the accuracy of the Wannier projection.
    (c, d) Topological edge states along [100] and [010] directions calculated from the
    Wannier-based tight-binding model, clearly showing the gapless helical edge states
    features of QSHI.}
    \label{fig:dft-soc}
\end{figure*}

Based on the relaxed structure, we compute the electronic band structures with and without spin-orbit coupling (SOC) to understand the role of SOC in determining the topological band inversions.
As shown in Fig.~\ref{fig:dft-soc}(a), in the absence of SOC, the valence and
conduction bands show band inversions at both $\Gamma$ and Y points, as the 
parity eigenvalues of the three relevant bands near the Fermi level are 
$(--+)$ at both momenta, indicating double band inversion. This configuration leads to trivial topological phase with $\mathbb{Z}_2$ invariant $\nu$=0.
Upon inclusion of SOC [Fig.~\ref{fig:dft-soc}(b)], the parity ordering at
the $\Gamma$ point changs to $(-,+,-)$, while the Y point remains unchanged.
This parity reordering removes the inversion at $\Gamma$, leaving only a 
single inversion at Y, and leads to a nontrivial $\mathbb{Z}_2$ topological
index $\nu_0$=1, confirming that monolayer TaIrTe$_4$ is a quantum spin Hall 
insulator (QSHI).

To provide an accurate Wannier representation of the DFT-calculated bands, we constructed maximally localized Wannier functions(MLWFs).
Specifically, we projected the Bloch wavefunctions onto the 
Ta ($s,p,d$ orbitals), Ir ($s$, $d$ orbitals), and Te($s$, $p$ orbitals),
totaling 124 orbitals. These orbitals well-capture the character of the 
states around the Fermi level, including all relevant electronic states within
the energy range of interest.
Among these 124 orbitals, 88 are occupied valence orbitals.
The resulting Wannier-interpolated band structure (red dashed lines)
exhibits excellent agreement with the DFT bands, especially near the Fermi
level, confirming the reliability and accuracy of our Wannier representation.

Furthermore, we calculate the topological edge states by iterative Green’s function method~\cite{sancho1985highly} along the [100] and [010] crystallographic directions.
As shown in Fig.~\ref{fig:dft-soc}(c,d), well-defined helical edge 
states emerge in both directions, traversing the bulk gap and 
confirming the nontrivial topology and bulk-edge correspondence of TaIrTe$_4$.
\subsection{Atom and Orbital Contributions to Electronic Bands}
\begin{figure*}[htb!] 
    \centering
    \includegraphics[width=0.8\textwidth]{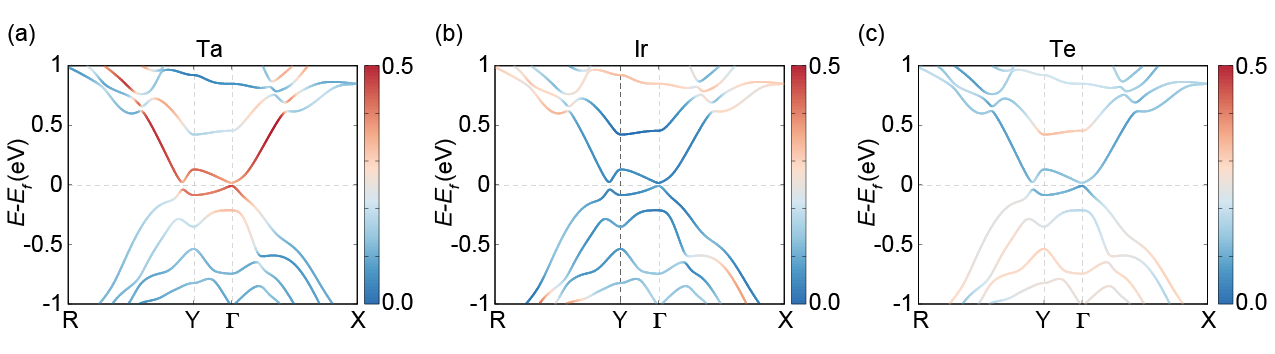}
    \caption{Atomic contribution to the electronic bands of TaIrTe$_4$.
    (a) Fat-band plot showing the dominant orbital contributions from Ta atoms.
    (b)-(c) Fat-band plots for Ir and Te atoms, respectively.
    The color scale represents the magnitude of atomic contribution,
    clearly indicating the states near the Fermi level are predominantly from Ta atoms.}
    \label{fig:dft-atom}
\end{figure*}

In the following, we analyze the atomic and orbital characteristics 
of the electronic bands near the Fermi level obtained from DFT 
calculations to prepare for the construction of the effective minimal 
tight-binding model.
Fig.~\ref{fig:dft-atom} shows the fat-band analysis including atomic 
contributions from Ta, Ir and Te atoms.
The results reveal that the electronic states around the Fermi level
are predominantly contributed by Ta atoms, as evidenced by the strong intensity near the Fermi level. Thus, the low-energy electronic physics of TaIrTe$_4$ is primarily dictated by the Ta atoms.
Therefore, we will focus on Ta atoms in the following analysis.

\begin{figure*}[htb!] 
    \centering
    \includegraphics[width=1.0\textwidth]{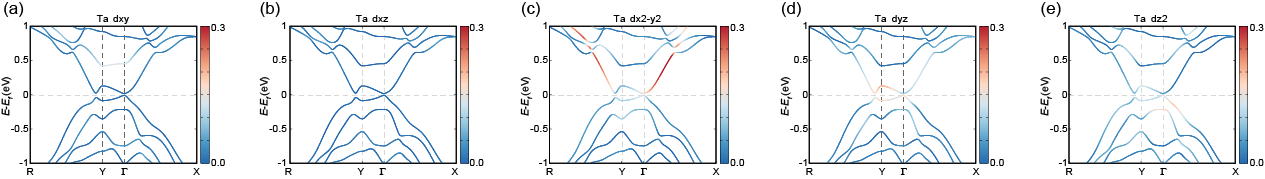}
    \caption{Orbital decomposition of Ta-5$d$ orbitals
    near the Fermi level.
    (a)-(e) Orbital-resolved fat-band plot for the Ta 5$d$ orbitals, including $d_{xy}$, $d_{xz}$, $d_{x^2-y^2}$, $d_{yz}$, $d_{z^2}$. The bands near the Fermi energy are dominated by the Ta $d_{x^2-y^2}$ and $d_{yz}$ orbitals, with minor contributions from $d_{z^2}$. The orbitals $d_{xy}$ and $d_{xz}$ have negligible contributions.}
    \label{fig:dft-orbital}
\end{figure*}

To further clarify the orbital character of the Ta-dominated bands 
near the Fermi level, we decomposed these bands into five Ta 5$d$ 
orbitals: $d_{xy}$, $d_{xz}$, $d_{z^2-y^2}$, $d_{yz}$, and $d_z^2$,
as shown in Fig.~\ref{fig:dft-orbital}(a)-(e).

The orbital decomposition clearly illustrates that the valence and conduction bands near the Fermi level mainly originate from the 
Ta $d_{x^2-y^2}$ [Fig.~\ref{fig:dft-orbital}(c)] and Ta $d_{yz}$ [Fig.~\ref{fig:dft-orbital}(d)] orbitals, with 
a smaller but noticeable contribution from the Ta $d_{z^2}$ orbital [Fig~\ref{fig:dft-orbital}(e)].
In contrast, the Ta $d_{xy}$ and $d_{xz}$ orbitals show essentially no
significant contribution to these bands [Fig.~\ref{fig:dft-orbital}(a)-(b)]. This indicates that the bands relevant to low-energy physics 
are predominately governed by the $d_{x^2-y^2}$ and $d_{yz}$ orbitals,
both of which exhibit significant dispersion and thus dominate the electronic structure near the Fermi level. Consequently, our minimal effective tight-binding model focuses primarily on these two active orbitals to accurately capture their dispersion characteristics.

\section{Computational Methods for $\mathbb{Z}_2$ and $\mathbb{Z}_4$ Invariants}\label{Ssec:Z2_Z4}

The characterization of topological phases in TaIrTe$_4$ requires a precise determination of its topological invariants. In this study, we compute both the $\mathbb{Z}_2$ and $\mathbb{Z}_4$ topological indices using Wilson loop methods and symmetry indicators based on density functional theory and tight-binding model calculations.

\subsection{Wilson Loop Calculation of the $\mathbb{Z}_2$ Invariant}\label{Ssec:Z2}
The Wilson-loop method~\cite{wang2019band,yu2011equivalent} is a wildly used method for topological classification of band structures. Here, we can define the Berry connection matrix for $\mathcal{N}$ occupied states for the small segment [$k_{x,i}$,$k_{x,i+1}$],
\begin{equation}
    F_{i,i+1}^{n,m}(k_y)=\langle u_n(k_{x,i},k_y)|u_m(k_{x,i+1},k_y)\rangle,
\end{equation}
where $u_{n,k_x,k_y}$ is the periodic part of Bloch function $\psi_{n,\mathbf{k}}$ and $k_{x,i}$=$\frac{2\pi i}{N_xa_x}$. 
The band indices $m$ and $n$ go over all occupied bands. 
Then, we can use a 
product of $F_{i,i+1}^{n,m}(k_y)$ to get the Berry phase matrix for a loop in discretized formula as
\begin{equation}
\mathcal{W}(k_y) = \prod_{i=0}^{N_x-1} F_{i,i+1}
\end{equation}
The Berry phase associated with the eigenvalues 
of above matrix, also known as Wilson bands, can be given by:
\begin{equation}
    \theta_n = -\mathrm{Im}\,\mathrm{log}(w_n)
\end{equation}
\noindent
where $w_n$ is the $n$th eigenvalue of Matrix $\mathcal{W}$ and $\theta_{n,k_y}$ is the Berry phase for the nth band along the $k_x$-loop for a fixed $k_y$.
The topological invariant can be obtained by the gauge-invariant winding number of the Wilson-loop evolution along $k_y$ for both the time-reversal broken and time-reversal invariant systems. A nontrivial $\mathbb{Z}_2$ invariant is characterized by a spectral flow of the Wilson bands across the Brillouin zone.

To compute $\mathbb{Z}_2$, we use the Fu-Kane formula, which expresses the strong TI index $\nu_0 \in \mathbb{Z}_2$ as:

\begin{equation}
(-1)^{\nu_0} = \prod_{K \in \text{TRIMs}} (+1)^{\frac{1}{2} n_K^+} (-1)^{\frac{1}{2} n_K^-} = (-1)^{\frac{1}{2} \sum_{K \in \text{TRIMs}} n_K^-},
\end{equation}
where $n_K^+$ ($n_K^-$) denotes the number of occupied bands with even (odd) parity at each time-reversal invariant momentum (TRIM) $K$. Due to Kramers pairing, $n_K^\pm$ is always even. This formulation enables an efficient determination of the topological phase.

\subsection{Symmetry Indicators and the $\mathbb{Z}_4$ Invariant}\label{Ssec:Z4}

To determine the higher-order topology of the system, we compute the $\mathbb{Z}_4$ invariant using symmetry indicators derived from the inversion eigenvalues at TRIMs. The $\mathbb{Z}_4$ invariant is defined as:
\begin{equation}
\kappa_1 \equiv \sum_{K \in \text{TRIMs}} (n_K^-)/2~~\textrm{mod}~~4 \in \mathbb{Z}.
\end{equation}
This index corresponds to the sum of the number of Kramers pairs with -1 parity eigenvalues of occupied bands at each TRIM point, modulo 4, distinguishing between different topological phases:
\begin{itemize}
\item $\kappa_1 = 0$: Trivial insulator.
\item $\kappa_1 = 1,3$: Topological insulator (TI) with $\nu_0 = 1$.
\item $\kappa_1 = 2$: Higher-order topological insulator (HOTI).
\end{itemize}

To extract the $\mathbb{Z}_4$ invariant, we utilize the eigenvalues of the inversion symmetry operator applied to the occupied bands at TRIMs, computed from the DFT-derived tight-binding Hamiltonian. This approach allows us to classify the topological character of different interaction-induced phases.
By applying these computational methods, we establish the topological phase diagram of TaIrTe$_4$ under different interaction and strain conditions, confirming the emergence of nontrivial QSHI, HOTI, and trivial insulating states.

\section{Eight-band Tight-binding Model}\label{Ssec:model}

To establish an effective low-energy Hamiltonian for TaIrTe$_4$, we start from the electronic structure analysis near the Fermi level. The DFT orbital decomposition indicates that the dominant orbitals contributing around the Fermi energy are $d_{x^2-y^2}$ and $d_{yz}$ from the two Ta atoms. Hence, we construct a minimal 8-band model based on these orbitals, incorporating spin degrees of freedom.

The general form of the 8-band Hamiltonian is given by:
\begin{equation}
H(\mathbf{k}) = H_0(\mathbf{k}) + H_{\text{SOC}}(\mathbf{k}),
\end{equation}
where $H_0(\mathbf{k})$ captures orbital hopping, and $H_{\text{SOC}}(\mathbf{k})$ introduces spin-orbit coupling (SOC).

\subsection{Optimized Wannier Fit (Hopping Terms)}
Considering crystal symmetries and nearest-neighbor hopping processes, the non-interacting Hamiltonian $H_0(\mathbf{k})$ is expressed as:
\begin{align}\label{eq:h0}
    H_0(\mathbf{k}) &= \left[ \mu_1 + t_1 \cos(a k_x) \right] \Gamma_1^{+} 
          + \left[ \mu_2 + d_1 \cos(a k_x) \right] \Gamma_1^{-} \nonumber \\
          &+ \left[(t_2 + t_3 e^{i b k_y})(1 + e^{i a k_x}) \right]\Gamma_2^{+} 
          + \left[(d_2 + d_3 e^{i b k_y})(1 + e^{i a k_x}) \right]\Gamma_2^{-} \nonumber \\
          &+ h_1 \cos(a k_x)\Gamma_3 
          + \left[(h_2 + h_3 e^{i b k_y})(1 + e^{i a k_x}) \right]\Gamma_4^{+} \nonumber \\
          &+ \left[(h_2 + h_3 e^{-i b k_y})(1 + e^{-i a k_x}) \right]\Gamma_4^{-} + h.c.,
\end{align}
where $t_{i}, d_{i}$ represent intra-orbital hoppings within $d_{x^2-y^2}$ and $d_{yz}$ orbitals, respectively, and $h_i$ describe inter-orbital couplings between these orbitals. Potentials for different orbitals are denoted by $\mu_1$ and $\mu_2$.

The Hamiltonian matrices $\Gamma_i$ are linear combinations of Pauli matrices acting in the orbital and sublattice basis:
\[
(a_{\mathbf{k}, A d_{x^2-y^2}s},\, a_{\mathbf{k}, B d_{x^2-y^2}s},\, a_{\mathbf{k}, A d_{yz}s},\, a_{\mathbf{k}, B d_{yz}s}),
\]
where $A,B$ denote two Ta atoms, and $s=\uparrow,\downarrow$ is the spin.

Explicitly, these matrices are defined as:
\begin{align*}
&\Gamma_1^{\pm} = \frac{1}{4}\tau_0(\sigma_0 \pm \sigma_3), &\Gamma_2^{\pm} &= \frac{1}{4}(\tau_1 + i\tau_2)(\sigma_0 \pm \sigma_3),\\
&\Gamma_3 = \frac{1}{2}\tau_0(\sigma_1 + i\sigma_2), &\Gamma_4^{\pm} &= \frac{1}{4}(\tau_1\pm i\tau_2)(\sigma_1+i\sigma_2),
\end{align*}
with $\sigma_i$ acting on orbital space, and $\tau_i$ on sublattice degrees of freedom.

\subsection{Spin-orbit Coupling (SOC) from Symmetry}
Considering symmetry constraints—translation symmetry $T(\mathbf{R})$, inversion symmetry $\mathcal{P}$, time-reversal symmetry $\Theta$, and glide mirror symmetry $\tilde{\mathcal{M}}_x=t(R_y/2)\mathcal{M}_x$—we introduce SOC into the Hamiltonian. In our chosen basis, these symmetries are represented by:
\begin{align*}
    \Theta &= is_y\tau_0\sigma_0\mathcal{K},\quad
    \mathcal{P} = s_0\tau_1\sigma_0,\quad
    \mathcal{M}_x = is_x\tau_0\sigma_0.
\end{align*}

These symmetry operations enforce constraints:
\begin{equation*}
OH(\mathbf{k})O^{-1} = H(\mathcal{O}\mathbf{k}),
\end{equation*}
where $O$ represents symmetry operations. The resulting symmetry-allowed SOC terms (up to nearest neighbors) are simplified as:
\begin{equation}\label{eq:hsoc}
H_{\text{SOC}}(\mathbf{k}) = \lambda_{1}\sin(a k_x) s_z\Gamma_5^{+} + \lambda_{2}\sin(a k_x) s_z\Gamma_5^{-},
\end{equation}
with the matrices:
\[
\Gamma_5^{\pm} = \frac{1}{4}\tau_3(\sigma_0 \pm \sigma_3).
\]
and parameters $\lambda_1$ and $\lambda_2$ quantify SOC strength for $d_{x^2-y^2}$ and $d_{yz}$ orbitals, respectively.

\begin{figure*}[ht] 
    \centering
    \includegraphics[width=0.9\textwidth]{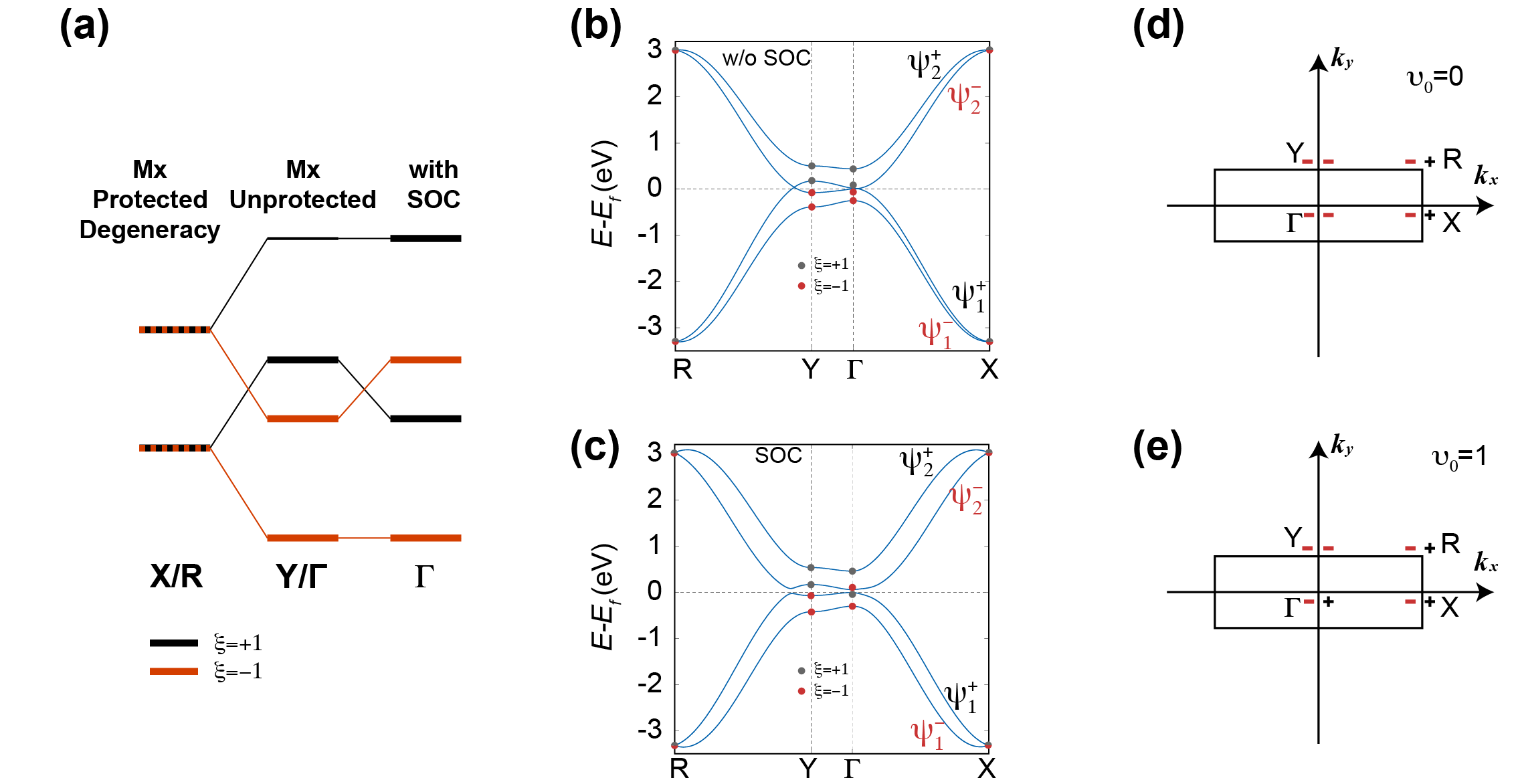}
    \caption{Mechanism of band inversion and parity eigenvalues from tight-binding model.(a) Schematic illustration of the band inversion mechanism, showing the evolution of bands under glide mirror symmetry ($\tilde{\mathcal{M}}_x$) protection, symmetry-breaking ($\tilde{\mathcal{M}}_x$ unprotected) and spin-orbit coupling (SOC). The black and orange lines denote states with positive ($\xi=+1$) and negative ($\xi=-1$) parity eigenvalues, respectively.
    (b,c) Band structures obtained from the tight-binding model (b) without SOC and (c) with SOC, respectively. Bands near the Fermi level are labeled as $\psi_1^\pm$ and $\psi_2^\pm$, and the parity eigenvalues at the time-reversal invariant momenta  are explicitly indicated by solid circles.
    (d,e) Schematic distribution of parity eigenvalues at the TRIM points ($\Gamma$, X, Y, R) without and with SOC, respectively. The parity eigenvalue switching at the $\Gamma$ point leads to a topological transition from a trivial insulator ($\nu_0=0$) to a quantum spin Hall insulator phase ($\nu_0=1$), indicating a single band inversion occurring at the Y point when SOC is included.
    }
    \label{fig:tb_model_s}
\end{figure*}

\subsection{Band structure and Parity Eigenvalues}\label{Ssec:tb-parity}
Figure~\ref{fig:tb_model_s} illustrates the band structures and parity eigenvalues ($\xi$) obtained from our optimized tight-binding model, elucidating the underlying band inversion mechanisms. Figure~\ref{fig:tb_model_s}(a) schematically depicts the evolution of bands under different symmetry conditions: initially, glide mirror symmetry ($\tilde{\mathcal{M}}_x$) protects the degeneracies at $k_x=\pi$ (X, R points). Without spin-orbit coupling (SOC), the degeneracies at X and R are robust due to the glide mirror symmetry, and these degenerate bands have opposite parity eigenvalues ($\xi=\pm 1$), originating from sublattice symmetry. However, at the $\Gamma$ and Y points, the absence of glide mirror protection leads to a splitting of bands originating from distinct orbitals. Consequently, the bands at these points undergo simultaneous parity inversion, resulting in identical parity configurations ($--+$), as explicitly shown in Fig.~\ref{fig:tb_model_s}(b). This scenario yields a trivial topological invariant ($\nu_0=0$), consistent with our density functional theory (DFT) calculations without SOC.

Upon incorporating SOC, the parity inversion at the $\Gamma$ point is reversed to a normal state, whereas the inversion at the Y point persists. Thus, as depicted in Fig.~\ref{fig:tb_model_s}(c), the parity eigenvalues at $\Gamma$ become ($-+-$), while at Y remain ($--+$). This single band inversion at the Y point gives rise to a nontrivial topological phase characterized by a $\mathbb{Z}_2$ invariant $\nu_0=1$. The resulting quantum spin Hall insulator phase is further confirmed by the presence of robust helical edge states along both [100] and [010] edges, as demonstrated in Fig.~\ref{fig:dft-soc} of Sec.~\ref{Ssec:dft} and Fig.~\ref{fig:fig1} in Sec.~\ref{mini-tb} of the main text.

Figures~\ref{fig:tb_model_s}(d-e) explicitly show the parity eigenvalue distributions at the four time-reversal invariant momenta, clearly highlighting the transition from a trivial to a nontrivial $\mathbb{Z}_2$ topological phase induced by SOC. Collectively, these results not only accurately reproduce the band eigenvalues near the Fermi level but also consistently capture the topology evolution revealed by our DFT calculations, validating the accuracy and robustness of our tight-binding model.
All fitted model parameters are summarized explicitly in Table~\ref{tbpars}.

\begin{table}[htbp]
\caption{Tight-binding model parameters of the eight-band model for monolayer TaIrTe$_4$. The model parameters correspond to Eqs.~\ref{eq:h0} and ~\ref{eq:hsoc}.}
\vspace{5pt} 
\centering
\resizebox{0.65\textwidth}{!}{%
\begin{tabular}{@{}c@{\hskip 20pt}ccc@{\hskip 20pt}c@{}}
\toprule
\textbf{Parameter} & \textbf{Value (meV)} & \hspace{1cm} & \textbf{Parameter} & \textbf{Value (meV)} \\
\midrule
$\mu_1$        & 1610        & & $h_{1}$        & -45     \\
$\mu_2$        & -1700    & & $h_{2}$        & 10  \\
$t_{1}$        & -700        & & $h_{3}$        & 10  \\
$t_{2}$        & 130     & & $\lambda_{1}$  & 200  \\
$t_{3}$        & -15        & & $\lambda_{2}$  & -150 \\
$d_{1}$        & 800      & & & \\
$d_{2}$        & 100     & & & \\
$d_{3}$        & -40         & & & \\
\bottomrule
\end{tabular}
}
\label{tbpars}
\end{table}

\subsection{Calculation of the Charge Susceptibility}

To analyze the electronic instabilities induced by Fermi surface nesting and vHS, we compute the static Lindhard charge susceptibility $\chi_0(\mathbf{q})$ based on the non-interacting 8-band tight-binding model. The susceptibility is evaluated using the following expression:

\begin{equation}
\chi_0(\mathbf{q}) = -\frac{1}{N_k} \sum_{\mathbf{k},n,m}
\frac{f(\varepsilon_{n\mathbf{k}}) - f(\varepsilon_{m\mathbf{k+q}})}{\varepsilon_{n\mathbf{k}} - \varepsilon_{m\mathbf{k+q}} + i\eta},
\end{equation}

where $f(\varepsilon)$ is the Fermi-Dirac distribution, $\varepsilon_{n\mathbf{k}}$ is the eigenvalue of the $n$-th band at momentum $\mathbf{k}$, and $\eta$ is a small broadening factor introduced to regularize the denominator. The summation runs over all band indices $n,m$ and discrete momentum points $\mathbf{k}$ in the Brillouin zone.

We performed this calculation on a dense $k$-mesh (typically $300\times300$) to ensure convergence near sharp features such as vHSs. The resulting $\chi_0(\mathbf{q})$ characterizes the momentum-resolved charge response of the system and can be used to identify possible nesting vectors or ordering wavevectors associated with CDW instabilities.

\section{Hartree-Fock Mean Field calculations}\label{Ssec:HFMF}
\subsection{Methods}\label{Ssec:HFMF-methods}
Here, we consider general Coulomb interactions in TaIrTe$_4$,
\begin{equation*}
    \hat{V} = \frac{1}{2}\iint d\mathbf{r}d\mathbf{r^{\prime}} 
    \sum_{s,s^\prime} \hat{c}_s^{\dagger}(\mathbf{r}) \hat{c}_{s^\prime}^{\dagger}(\mathbf{r^\prime})
    V_{\mathrm{int}}(|\mathbf{r}-\mathbf{r^\prime}|)
    \hat{c}_{s^\prime}(\mathbf{r^\prime}) \hat{c}_s(\mathbf{r})
\end{equation*}
where $\hat{c}_s(\mathbf{r})$ is electron annihilation operator at continuous coordinate $\mathbf{r}$ with spin $s$.
It can be projected to the Wannier lattice basis as:
\begin{equation}
    \hat{c}_{s}^{\dagger}(\mathbf{r}) = \sum_{\mathbf{R},\alpha} w_{\mathbf{R},\alpha}^*(\mathbf{r})a_{\mathbf{R},\alpha s}^\dagger
\end{equation}
where subscript $\mathbf{R}$ labels a Bravis lattice site, 
$\alpha$ is a sublattice/orbital index, 
and $s$ is a spin index.
The interaction can be expressed as:
\begin{equation}
    \hat{V}_{\mathrm{int}}=
    \frac{1}{2}\sum_{\mathbf{R_1}\mathbf{R_2}\mathbf{R_3}\mathbf{R_4}} \sum_{\alpha \beta \beta' \alpha'}\sum_{ss^\prime}
    \hat{a}_{\mathbf{R_1},\alpha s}^\dagger \hat{a}_{\mathbf{R_2},\beta s'}^{\dagger}
    V_{\mathbf{R_1}\mathbf{R_4},\mathbf{R_2}\mathbf{R_3}}^{\alpha\alpha' s, \beta\beta's'}
        \hat{a}_{\mathbf{R_3},\beta' s'}
        \hat{a}_{\mathbf{R_4},\alpha' s}
\end{equation}
where
\begin{equation}
    V_{\mathbf{R_1}\mathbf{R_4},\mathbf{R_2}\mathbf{R_3}}^{\alpha\alpha' s, \beta\beta's'}
    =\iint d\mathbf{r}d\mathbf{r'}
    V_{\mathrm{int}}(|\mathbf{r-r'}|)w_{\mathbf{R_1},\alpha}^*(\mathbf{r})w_{\mathbf{R_2},\beta}^*(\mathbf{r}')w_{\mathbf{R_3},\beta'}(\mathbf{r'})w_{\mathbf{R_4},\alpha'}(\mathbf{r})
\end{equation}
Here, we assume that the ``density-density" interaction takes an important role in the system,
$V_{\mathbf{R_1}\mathbf{R_4},\mathbf{R_2}\mathbf{R_3}}^{\alpha\alpha' s, \beta\beta's'}\approx V_{\mathbf{R_1}\mathbf{R_1},\mathbf{R_2}\mathbf{R_2}}^{\alpha\alpha s,\beta\beta s'}\equiv V_{\mathbf{R_1}\alpha s,\mathbf{R_2}\beta s'}$.
Then the Coulomb interaction can be written as:
\begin{align}
    \hat{V}_{\mathrm{int}}=&
    \frac{1}{2}\sum_{\mathbf{R_1}\mathbf{R_2}} \sum_{\alpha \beta}\sum_{ss^\prime}
    \hat{a}_{\mathbf{R_1},\alpha s}^\dagger \hat{a}_{\mathbf{R_2},\beta s'}^{\dagger}
V_{\mathbf{R_1},\alpha s,\mathbf{R_2}\beta s'}
        \hat{a}_{\mathbf{R_2},\beta s'}
        \hat{a}_{\mathbf{R_1},\alpha s}\nonumber \\
        =&\frac{1}{2}\sum_{\mathbf{R_1}\alpha\neq\mathbf{R_2}\beta}
        \sum_{ss^\prime}
    \hat{a}_{\mathbf{R_1},\alpha s}^\dagger \hat{a}_{\mathbf{R_2},\beta s'}^{\dagger}
V_{\mathbf{R_1}\alpha,\mathbf{R_2}\beta}
        \hat{a}_{\mathbf{R_2},\beta s'}
        \hat{a}_{\mathbf{R_1},\alpha s} \nonumber \\
        &+\sum_{\mathbf{R}\alpha} U\hat{a}^\dagger_{\mathbf{R},\alpha\uparrow}\hat{a}^\dagger_{\mathbf{R},\alpha\downarrow}a_{\mathbf{R},\alpha\downarrow}a_{\mathbf{R},\alpha\uparrow}
\end{align}
where we separate off the on-site Hubbard term with coefficient $U$. In our approximation, the hopping process of electrons getting through the lattice is modified by the Coulomb interaction. In our main text, we focus on the density-density interaction as:
\begin{align}
    \hat{H}_{\mathrm{int}}
        =&U\sum_{\mathbf{R}\alpha} \hat{n}_{\mathbf{R},\alpha\uparrow}\hat{n}_{\mathbf{R},\alpha\downarrow}+\frac{1}{2}\sum_{\mathbf{R_1}\alpha\neq\mathbf{R_2}\beta}^{3rd NN}
        \sum_{ss^\prime}V_{\mathbf{R_1}\alpha,\mathbf{R_2}\beta} \hat{n}_{\mathbf{R_1,\alpha s}}\hat{n}_{\mathbf{R_2},\beta s'}~.
\end{align}
where $\hat{n}$ is the density operator, $U$ is the strength of the
onsite Hubbard interaction, and $V$ is the amplitude of the density-density interaction between sites.

We use a double-gate method to model screening effects and get the values of U and V. A general Coulomb interaction V(r) is screened by a pair of metallic gates, each a distance $\pm d$ from the 2D TaIrTe$_4$. The gates are assumed to be perfect conductors so that the screened Coulomb interaction can be expressed as
\begin{equation}
    V(r) = \frac{e^2}{ 4\pi\epsilon_0\epsilon_r}\sum_{z\in\mathbb{Z}}
    \frac{(-1)^z}{\sqrt{r^2+(2dz)^2}}
\end{equation}
where $\epsilon_r$ is a background dielectric constant and the 
distance between two gates is $d=100$ \AA. We have checked the influence of the distance $d$ for our results. The HFMF phase diagram of $d$=200 \AA \ is similar to $d$=100 \AA.
To simulate the onsite screening, we
vary U by considering r in a range of 0.8 \AA ~to 2.0 \AA.

For the HFMF loop, we again write the extended Hubbard Hamiltonian defined on a tight-binding basis:
\begin{align*}\hat{H} 
        &= \hat{H}_0+\hat{H}_{\mathrm{int}}\\
        &=\sum_{\mathbf{R_1}, \mathbf{R_2}} \sum_{\alpha \beta s} t_{\substack{\mathbf{R_1 - R_2},\alpha \beta}} a^{\dagger}_{\mathbf{R_1},\alpha s}a_{\mathbf{R_2},\beta s} \\
&+ U\sum_{\substack{\mathbf{R,\alpha}}}
\hat{n}_{\mathbf{R},\alpha\uparrow}\hat{n}_{\mathbf{R},\alpha\downarrow} \\
&+ 
\frac{1}{2} \sum_{\substack{(\mathbf{R_1}\alpha)\neq(\mathbf{R_2}\beta)}}\sum_{ss'}
V_{\mathbf{R_1\alpha},\mathbf{R_2\beta}} \hat{n}_{\mathbf{R_1},\alpha s} \hat{n}_{\mathbf{R_2},\beta s'}\end{align*}
where $\mathbf{R_1}$ and $\mathbf{R_2}$ label a lattice sites,  $\alpha, \beta$ are composite sublattice-orbital indices, and $s, s'$ are spin indices. The matrix of correlation functions of a generic bilinear Hamiltonian are obtained by eigendecomposition:
\begin{align}
&\mathbf{c}^{\dagger} H \mathbf{c} = \mathbf{c}^{\dagger} U \Lambda U^{\dagger} \mathbf{c} = \boldsymbol{\gamma}^{\dagger} \Lambda \boldsymbol{\gamma} \nonumber \\
&\Rightarrow \langle (\mathbf{c}^{\dagger})^{T} (\mathbf{c}^{})^{T}\rangle = \langle (\boldsymbol{\gamma}^{\dagger} U^{\dagger})^{T} (U \boldsymbol{\gamma})^{T}\rangle = \langle U^{*} (\boldsymbol{\gamma}^{\dagger})^{T}  \boldsymbol{\gamma}^{T} U^{T}\rangle = U^{*} \langle (\boldsymbol{\gamma}^{\dagger})^{T}  \boldsymbol{\gamma}^{T} \rangle U^{T} = U^{*} \textrm{diag}(n_{F}(\lambda_i)) U^{T}
\end{align}
where $\mathbf{c}$ is a vector of annihilation operators. The Hartree-Fock approximation of the density-density interaction term yields such a bilinear:
\begin{align*}
\hat{n}_{\mathbf{R_1},\alpha s} \hat{n}_{\mathbf{R_2},\beta s'} 
&= a^{\dagger}_{\mathbf{R_1},\alpha s} a_{\mathbf{R_1},\alpha s} 
   a^{\dagger}_{\mathbf{R_2},\beta s'} a_{\mathbf{R_2},\beta s'} \\
&\approx \langle a^{\dagger}_{\mathbf{R_1},\alpha s} a_{\mathbf{R_1},\alpha s}\rangle 
                a^{\dagger}_{\mathbf{R_2},\beta s'}a_{\mathbf{R_2},\beta s'} +  \langle 
                a^{\dagger}_{\mathbf{R_2},\beta s'}a_{\mathbf{R_2},\beta s'} \rangle a^{\dagger}_{\mathbf{R_1},\alpha s} a_{\mathbf{R_1},\alpha s} && \textrm{Hartree} \\
&- \langle a^{\dagger}_{\mathbf{R_1},\alpha s} a_{\mathbf{R_2},\beta s'}  \rangle  
            a^{\dagger}_{\mathbf{R_2},\beta s'} a_{\mathbf{R_1},\alpha s} - \langle  
            a^{\dagger}_{\mathbf{R_2},\beta s'} a_{\mathbf{R_1},\alpha s} \rangle a^{\dagger}_{\mathbf{R_1},\alpha s} a_{\mathbf{R_2},\beta s'}  && \textrm{Fock} \\
\end{align*}
We demand translational invariance by picking a preferred origin and forcing all other correlation functions to match, i.e.
\begin{equation} \langle a^{\dagger}_{\mathbf{R_1},\alpha s} a_{\mathbf{R_2},\beta s'}  \rangle = \langle a^{\dagger}_{\mathbf{R_1 - R_2},\alpha s} a_{\mathbf{0},\beta s'}  \rangle \end{equation}
which means the correlation functions depend only on the difference in coordinates. This yields a new, also translationally invariant, bilinear Hamiltonian:
\begin{align*}\hat{H}_{HF} = 
\sum_{\substack{\mathbf{R_1}\mathbf{R_2},\\\alpha \beta s s'}} h_{\mathbf{R_1 - R_2},\alpha s \beta s'} a^{\dagger}_{\mathbf{R_1},\alpha s}a_{\mathbf{R_2},\beta s'}\end{align*}
with
\begin{align*}
h_{\mathbf{R_1}-\mathbf{R_2},\alpha s \beta s'} &= 
t_{\mathbf{R_1-R_2},\alpha \beta}\delta_{ss'}  \\
&+ U \delta_{\mathbf{R_1}-\mathbf{R_2}, \mathbf{0}}\delta_{\alpha \beta} \begin{bmatrix} 
\langle a^{\dagger}_{\mathbf{R_1}\alpha \downarrow} a^{}_{\mathbf{R_1}\alpha \downarrow} \rangle & 
-\langle a^{\dagger}_{\mathbf{R_1}\alpha \downarrow} a^{}_{\mathbf{R_1}\alpha \uparrow} \rangle \\ 
-\langle a^{\dagger}_{\mathbf{R_1}\alpha \uparrow} a^{}_{\mathbf{R_1}\alpha \downarrow} \rangle & 
\langle a^{\dagger}_{\mathbf{R_1}\alpha \uparrow} a^{}_{\mathbf{R_1}\alpha \uparrow} \rangle \end{bmatrix}_{ss'}   \\
&+ \delta_{\mathbf{R_1}-\mathbf{R_2},\mathbf{0}} \delta_{\alpha\beta} \delta_{ss'} \sum_{\mathbf{R} \neq \mathbf{0},\gamma, s''} V_{\mathbf{0\alpha},\mathbf{R\gamma}}  
\langle a^{\dagger}_{\mathbf{R}\gamma s''} a^{}_{\mathbf{R}\gamma s''} \rangle \\
&- V_{\mathbf{R_1\alpha},\mathbf{R_2\beta}} \langle a^{\dagger}_{\mathbf{R_2}\beta s'} a^{}_{\mathbf{R_1}\alpha s} \rangle 
\end{align*}
which leads to a fixed-point equation for the correlation functions. Our relatively large Hartree-Fock problem is tractable for the following reasons. First, translational invariance simplifies the calculation of correlation functions. The Fourier transform of $H_{HF}$ is block diagonal in $\mathbf{k}$:
\begin{align*}
\hat{H}_{HF} &= \frac{1}{N_{k}}\sum_{\mathbf{k}}\left[\sum_{\mathbf{R_1 - R_2}} h_{\mathbf{R_1 - R_2},\alpha s \beta s'} e^{-i\mathbf{k}\cdot(\mathbf{R_1 - R_2})}  \right] a^{\dagger}_{\mathbf{k},\alpha s}a^{}_{\mathbf{k},\beta s'} \\
&= \frac{1}{N_{k}} \sum_{\mathbf{k}} h_{\mathbf{k},\alpha s \beta s'} a^{\dagger}_{\mathbf{k},\alpha s}a^{}_{\mathbf{k},\beta s'} 
\end{align*}
yielding correlation functions $\langle c^{\dagger}_{\mathbf{k},\alpha s}c_{\mathbf{k},\beta s'} \rangle$ by the above derivation. $N_k$ is the number of points in the $k$-space mesh. The small block matrices can be diagonalized in parallel on a GPU. The real space correlation functions are of course given by the inverse transform:
\begin{align*}
\langle a^{\dagger}_{\mathbf{R_1},\alpha s} a^{}_{\mathbf{R_2},\beta s'} \rangle = \frac{1}{N_{k}} \sum_{\mathbf{k}} e^{i\mathbf{k}\cdot(\mathbf{R_1 - R_2})}\langle a^{\dagger}_{\mathbf{k}, \alpha s} a^{}_{\mathbf{k}, \beta s'} \rangle
\end{align*}
which are plugged back into the real-space Hamiltonian to construct the next Hartree-Fock iterate. The second acceleration is to use a direct quasi-Newton method \cite{Zhou2011aquasi}, which allows us to converge the large Hartree-Fock fixed point equations, converging 120 orbital problems in relatively few iterations.

\subsection{Topological Properties of Dual QSHI Phase}\label{Ssec:HFMF-results-dual-QSHI}
\begin{figure}[htbp] 
    \centering
    \includegraphics[width=0.8\textwidth]{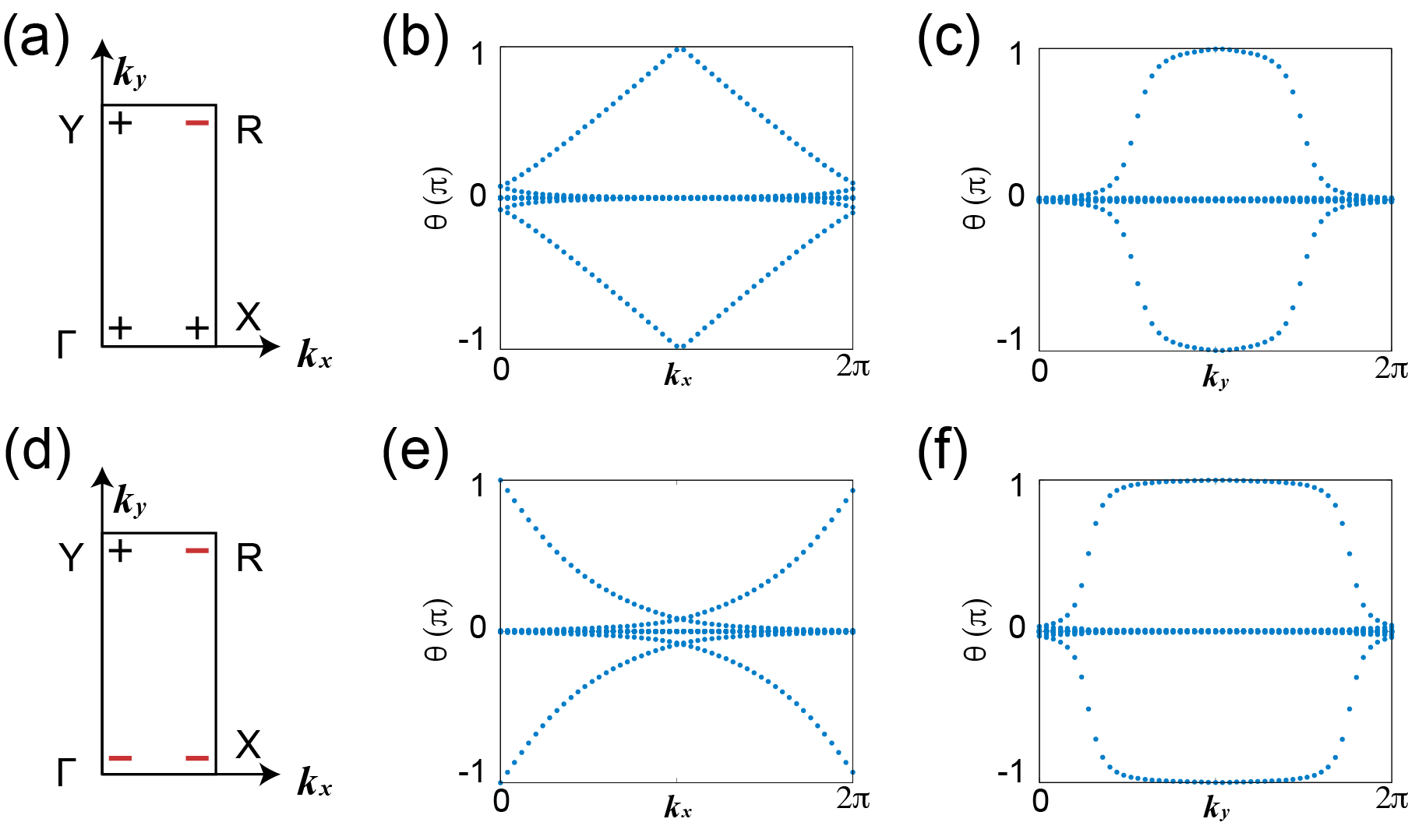}
    \caption{Topological properties for the dual QSHI phase at $U/V_1=2.25$, $\epsilon=13$.
    (a)-(c) Parity eigenvalues and Wilson loop evolutions along $k_x$ and $k_y$ for the $n_e$ gap, indicating a nontrivial topology with band inversion at the R point.
    (d)-(f) Corresponding results for the CNP gap, demonstrating the identical topological nature and confirming the dual QSHI state.}
    \label{fig:dual_qshi}
\end{figure}

Figure~\ref{fig:dual_qshi} provides supplementary details of the topological properties associated with the dual QSHI phase discussed in the main text. In this phase, both the $n_e$ gap (between the 62nd and 63rd bands) and the CNP gap (between the 60th and 61st bands) exhibit identical nontrivial $\mathbb{Z}_2$ topological invariants ($\nu_0=1$). Specifically, at the representative parameter point $U/V_1=2.25$, $\epsilon=13$, the correlated insulating state belongs to the dual QSHI phase. 
The band inversions for $n_e$ gap occurs at the R point in momentum space, as indicated by the parity product distributions in panels (a). 
In contrast, for the CNP gap [Fig.~\ref{fig:dual_qshi}(d)], the band inversion occurs at the Y point, as reflected in the parity configuration.
The topological character of each gap is further confirmed by the evolution of the Wannier centers. 
For the $n_e$ gap (panels (b) and (c)), the Wannier center evolution shows nontrivial windings along both $k_x$ and $k_y$ directions, characterized by phase jumps at $\pm\pi$, indicating clearly defined QSHI behavior. Similar topological features are observed at the CNP gap, as shown in panels (e) and (f).  
Despite the difference in inversion momenta, both gaps yield a nontrivial $\mathbb{Z}_2$ gap, constituting the dual QSHI phase. 
These features also explain the presence of helical edge states in both gaps, as described in the main text. These findings fully reproduce and support the existence of the dual QSHI state, confirming that electron interactions stabilize robust topological insulating phases at both $n_e$ and CNP gaps.

\subsection{Topological Properties of the \textbf{Dual HOTI} Phase}\label{Ssec:HFMF-results-dual-HOTI}
\begin{figure}[htb!]
    \centering
    \includegraphics[width=0.9\linewidth]{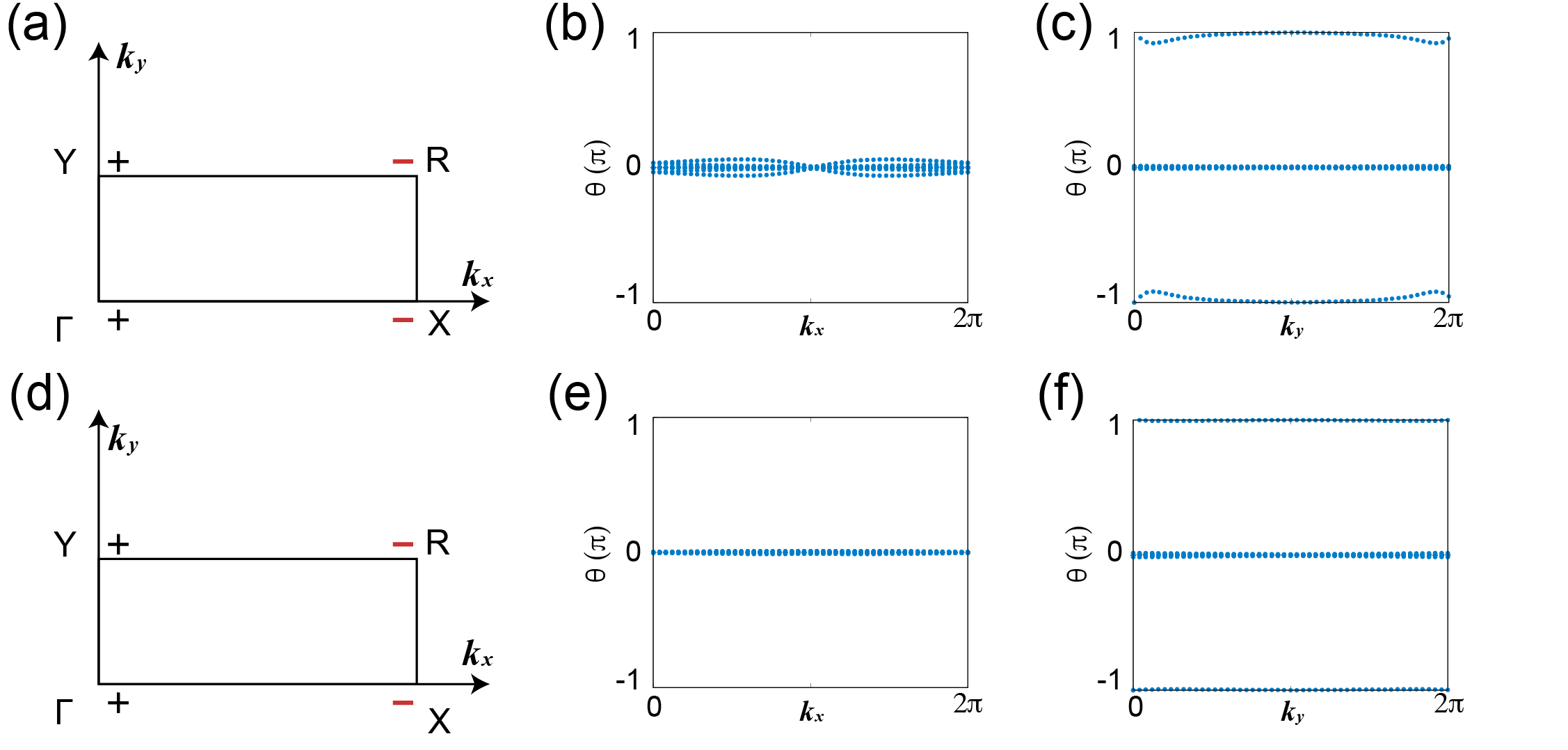}
    \caption{
    Topological properties for the dual HOTI phase.
    (a,d) Parity products at the four time-reversal invariant momenta for the occupied bands at the $n_e$ gap (a) and the CNP gap (d), both showing the same pattern $(+,-,+,-)$, indicating a $\mathbb{Z}_2$ index $\nu = 0$ and a $\mathbb{Z}_4$ invariant $\kappa_1 = 2$.
    (b,e) Wilson loop spectra along the $k_x$ direction for the $n_e$ and CNP gaps, respectively, both showing trivial winding (flat phases).
    (c,f) Wilson loop spectra along the $k_y$ direction for the $n_e$ and CNP gaps, respectively, showing two branches pinned at $\pm \pi$, confirming the higher-order topology.
    }
    \label{fig:dual_HOTI}
\end{figure}

In Fig.~\ref{fig:dual_HOTI}, we further characterize the topological properties of the dual HOTI phase. At both the $n_e$ gap and CNP gap, the parity product distributions of the occupied bands are identical, as shown in Fig.~\ref{fig:dual_HOTI}. Specifically, the parity eigenvalues at the four TRIMs follow the pattern $(+,-,+,-)$, leading to a $\mathbb{Z}_2$ topological index $\nu_0 = 0$. However, the system exhibits a higher-order topology captured by the $\mathbb{Z}_4$ invariant $\kappa_1 = 2$.

The Wilson loop spectra further confirm this higher-order topological nature. Along the \(k_x\) direction, the Wilson loop phases remain zero across the Brillouin zone, indicating a trivial winding [Fig.~\ref{fig:dual_HOTI}(b,e)]. In contrast, along the \(k_y\) direction, the Wilson loop exhibits two branches pinned at \(\pm \pi\), as shown in Fig.~\ref{fig:dual_HOTI}(c,f). This behavior is consistent with a quantized polarization localized at the edges, confirming the presence of higher-order topology in both the $n_e$ and CNP gaps. 
These topological features directly lead to the emergence of corner states, as observed in the main text, further validating the existence of the dual HOTI phase.

\begin{figure}[htb!]
    \centering
    \includegraphics[width=0.8\linewidth]{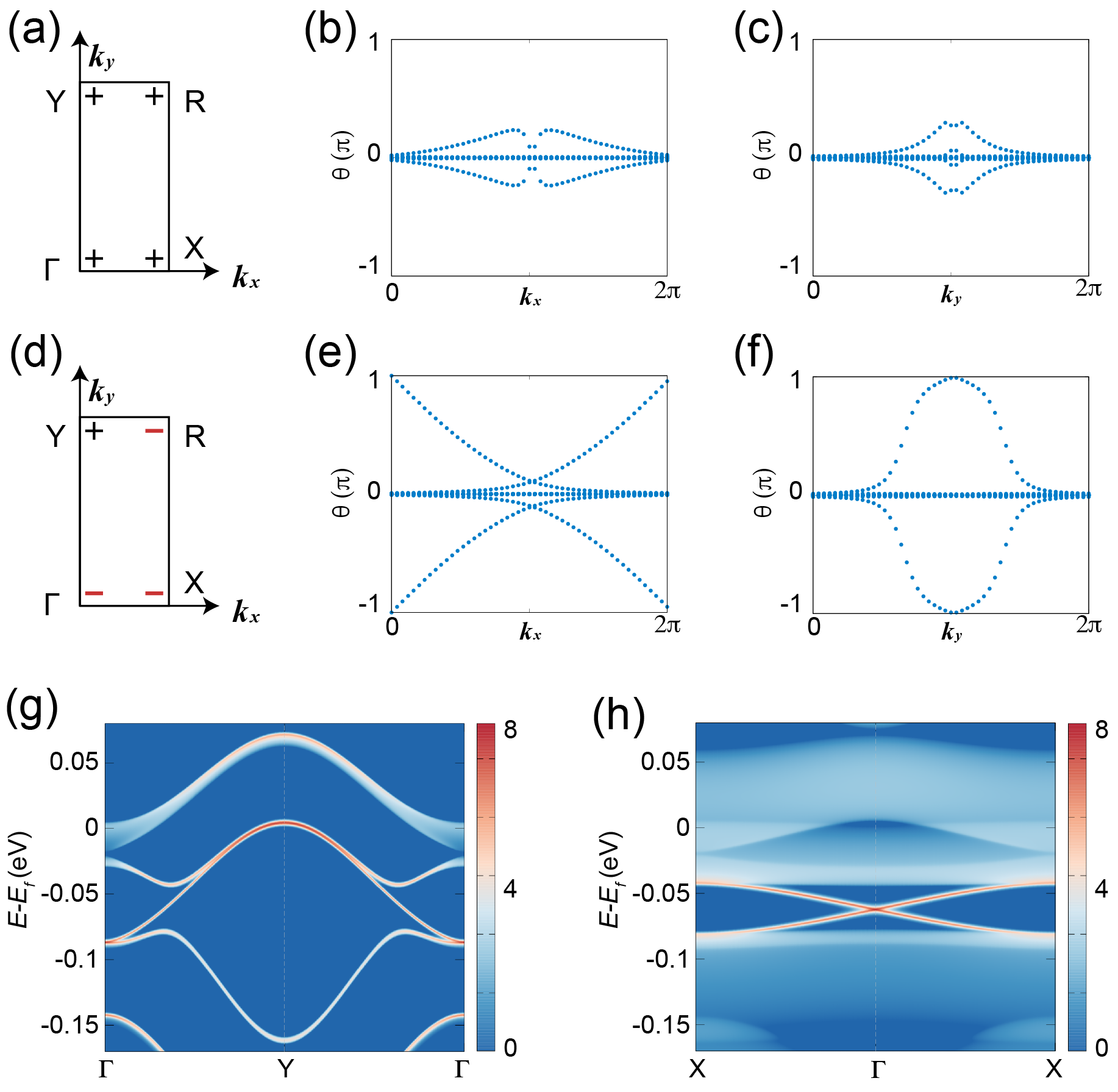}
    \caption{
    Topological properties for the QSHI+Metal phase.
    (a,d) Parity eigenvalues of the occupied bands at the $n_e$ gap (a) and the CNP gap (d), respectively. The $n_e$ bands have $\nu_0 = 0$ with trivial parity configuration, while the CNP bands exhibit a nontrivial $\nu_0 = 1$ index due to band inversion at the Y point.
    (b,e) Wilson loop along $k_x$ direction for the $n_e$ and CNP bands. Flat phases indicate no winding for the $n_e$ bands, while the CNP bands exhibit a nontrivial flow.
    (c,f) Wilson loop along $k_y$, confirming the $\mathbb{Z}_2$ index change only in the CNP gap.
    (g,h) Edge spectra along [010] and [100] directions. Helical edge states appear only at the CNP gap (h), while no clear topological edge modes are found at the $n_e$ gap (g), due to the absence of a nontrivial gap.
    }
    \label{fig:metal_qshi}
\end{figure}

\subsection{Topological Characterization of the QSHI+Metal Phase}\label{Ssec:HFMF-results-metal-HOTI}
In the regime of weak electron-electron interactions, particularly at large dielectric constants (e.g., $\epsilon = 20$), our Hartree-Fock mean-field calculations reveal a distinct phase in which the system exhibits a metallic state at the $n_e$ gap while preserving a nontrivial topological insulator at the charge neutrality point. This coexistence defines the QSHI+metal phase.

As shown in Fig.~\ref{fig:metal_qshi}(a–c), the $n_e$ gap does not develop a global insulating gap across the Brillouin zone. Instead, it remains partially gapless, indicating a metallic character. Nevertheless, we can still compute the parity eigenvalues of the occupied states at the four time-reversal invariant momenta, which yield a trivial configuration $\nu_0 = 0$. This is further supported by the Wilson loop calculations along both $k_x$ and $k_y$ directions [Fig.~\ref{fig:metal_qshi}(b,c)], where no winding or flow in the Wannier phases is observed. This metallic character stems from insufficient electron correlations, as strong dielectric screening effectively weakens both onsite and intersite interactions. Consequently, the system remains in a Fermi-liquid-like state near the doping level, with no topological protection.

In contrast, the CNP gap remains robust and topologically nontrivial, even in the presence of reduced interactions. As shown in Fig.~\ref{fig:metal_qshi}(d–f), the parity products indicate a band inversion occurring at the Y point, leading to a nontrivial $\mathbb{Z}_2$ index $\nu_0 = 1$. This conclusion is verified by the Wilson loop spectra, where a characteristic gapless flow emerges in the $k_x$ direction [Fig.~\ref{fig:metal_qshi}(e)], confirming the persistence of a QSHI state at charge neutrality.

The topological distinction between these two energy gaps is further revealed by the calculated edge states. As shown in Fig.~S8(g)-(h), the CNP gap supports well-defined helical edge states connecting conduction and valence bands, consistent with the presence of a QSHI phase. In contrast, no clear edge states are observed within the $n_e$ gap region, demonstrating the absence of a full gap and a topologically trivial metallic character.

These results establish that under weak interaction conditions, the system undergoes a topological differentiation between the $n_e$ and CNP gaps: the $n_e$ gap becomes metallic, while the CNP retains a correlated QSHI character. This QSHI + Metal phase highlights the critical role of interaction strength in stabilizing different topological phases and provides a concrete example of gap-selective topological behavior within a single correlated system.

\section{Strain effect}\label{Ssec:strain_effect}
\subsection{Topological Properties of QSHI under Compressive Strain}

\begin{figure}[htb!]
    \centering
    \includegraphics[width=0.75\textwidth]{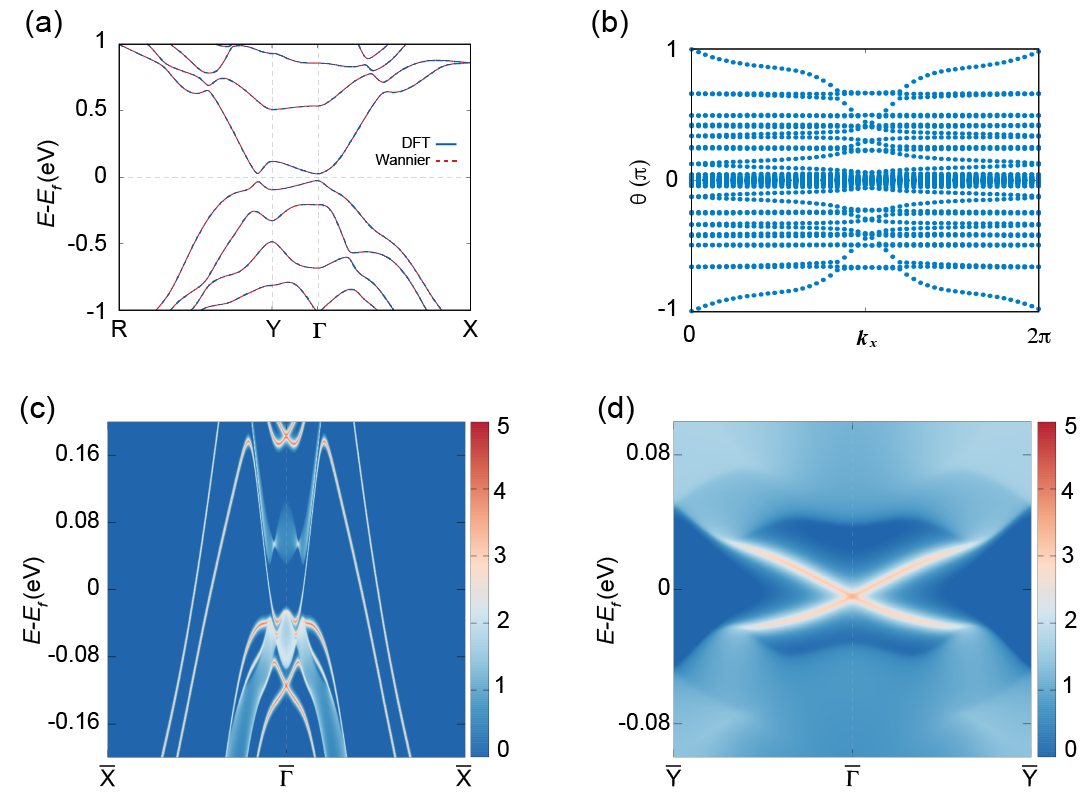}
    \caption{Topological properties for QSHI under -1\% strain.
    (a) DFT band structure (blue) and Wannier interpolation (red dashed) under $-1\%$ uniaxial strain. The band gap increases to approximately 38 meV.
    (b) Wilson loop evolution of the occupied bands along the $k_x$ direction, showing a nontrivial winding with $\pm\pi$ phase jump at $k_x = 0$, confirming the $\mathbb{Z}_2$ invariant $\nu_0 = 1$.
    (c) Topological edge states along the [100] direction, showing gapless helical edge modes.
    (d) Topological edge states along the [010] direction, also showing helical edge modes protected by time-reversal symmetry.}
    \label{fig:S10}
\end{figure}

To confirm the strain-induced topological response in monolayer TaIrTe$_4$, we perform first-principles calculations under $-1\%$ uniaxial compressive strain. As shown in Fig.~\ref{fig:S10}(a), the DFT-calculated band structure (blue) and Wannier-interpolated bands (red dashed) match well, revealing a significantly enhanced energy gap of approximately 38 meV. This enlarged gap is advantageous for observing robust edge-state transport in experiments.

We further evaluate the topological nature of the system by computing the parity eigenvalues of the occupied bands at the four time-reversal invariant momenta: $\Gamma$, X, Y, and R. The resulting parity products are $(+,+,-,+)$, which match the unstrained case and confirm that the system remains in the quantum spin Hall insulator phase under compressive strain.

Figure~\ref{fig:S10}(b) presents the Wilson loop evolution of the Wannier centers along the $k_x$ direction. The Wannier phases exhibit a gapless flow and cross $\theta = \pm\pi$ at $k_x = 0$, indicating a nontrivial $\mathbb{Z}_2$ index $\nu_0 = 1$. This agrees well with the parity product analysis. Figures~\ref{fig:S10}(c) and (d) show the edge state spectra along the [100] and [010] directions, respectively. In both cases, a pair of gapless helical edge modes traverse the bulk gap, further confirming the nontrivial topological character. These edge states are protected by time-reversal symmetry and are a hallmark of QSHI.

In summary, our results demonstrate that the QSHI phase in TaIrTe$_4$ is robust against moderate compressive strain. The enlarged band gap improves the experimental accessibility of the edge states, offering a promising route for realizing strain-enhanced topological transport in correlated 2D materials.

\begin{figure}[htb!]
    \centering
    \includegraphics[width=1.0\linewidth]{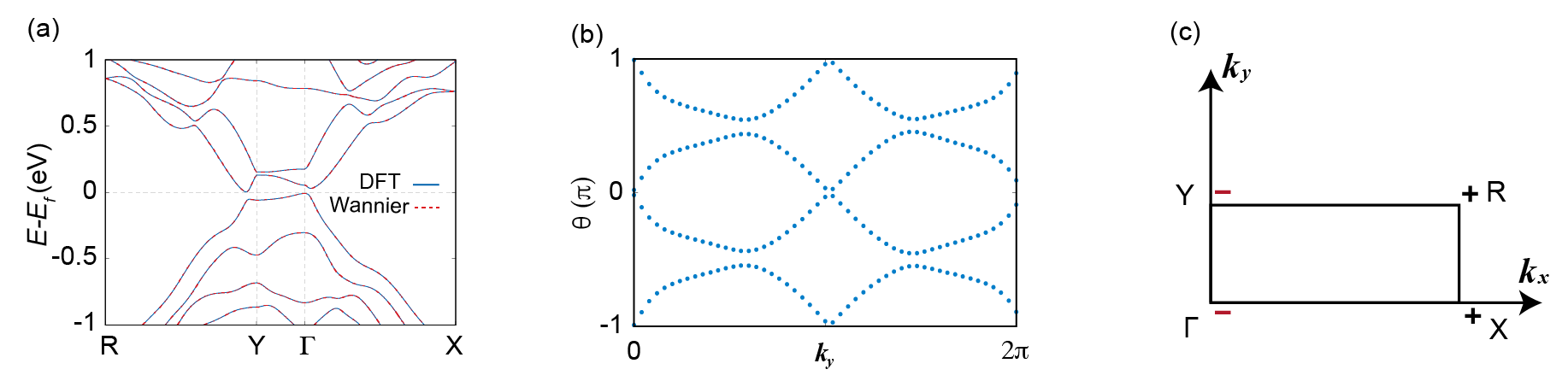}
    \caption{Topological properties of TaIrTe$_4$ under $+5\%$ tensile strain.
    (a) DFT band structure (blue) and Wannier interpolation (red dashed) under strain, showing a new band inversion at the $\Gamma$ point.
    (b) Wilson loop evolution along the $k_y$ direction. The appearance of $\pm\pi$ flow signals higher-order topology.
    (c) Parity products at the four TRIM points ($\Gamma$, X, Y, R), yielding the pattern $(-,+,-,+)$. This corresponds to a trivial $\mathbb{Z}_2$ index $\nu_0 = 0$ and nontrivial $\mathbb{Z}_4$ index $\kappa_1 = 2$, indicative of a HOTI phase.
    }
    \label{fig:S11}
\end{figure}

\subsection{Topological Properties under Tensile Strain}\label{Ssec:strain_effect_dft}

To illustrate the strain-driven topological phase transition from a QSHI to a HOTI phase, we perform first-principles calculations at $+5\%$ tensile strain. As shown in Fig.~\ref{fig:S11}(a), the DFT and Wannier-interpolated bands reveal a clear band inversion at the $\Gamma$ point, which is absent under zero strain. This inversion accompanies the one already present at Y, leading to a double band inversion scenario.

The Wilson loop calculation along the $k_y$ direction [Fig.~\ref{fig:S11}(b)] exhibits two Wannier bands crossing $\theta = \pm\pi$, indicating a nontrivial higher-order topology. Consistently, the parity products at the four time-reversal invariant momenta $(\Gamma, X, Y, R)$ are $(-,+,-,+)$ [Fig.~\ref{fig:S11}(c)], confirming that the system’s $\mathbb{Z}_2$ invariant becomes $\nu_0 = 0$, while the $\mathbb{Z}_4$ invariant $\kappa_1 = 2$, characteristic of a HOTI phase.

These results validate our main-text phase diagram and demonstrate that under sufficient tensile strain, the system transitions into a HOTI state with double parity inversions and boundary-localized topological corner modes.

\subsection{\textit{QSHI + HOTI} phase under 2\% tensile strain
}\label{Ssec:strain_effect_hfmf_1}

\begin{figure}[htb!]
    \centering
    \includegraphics[width=0.9\linewidth]{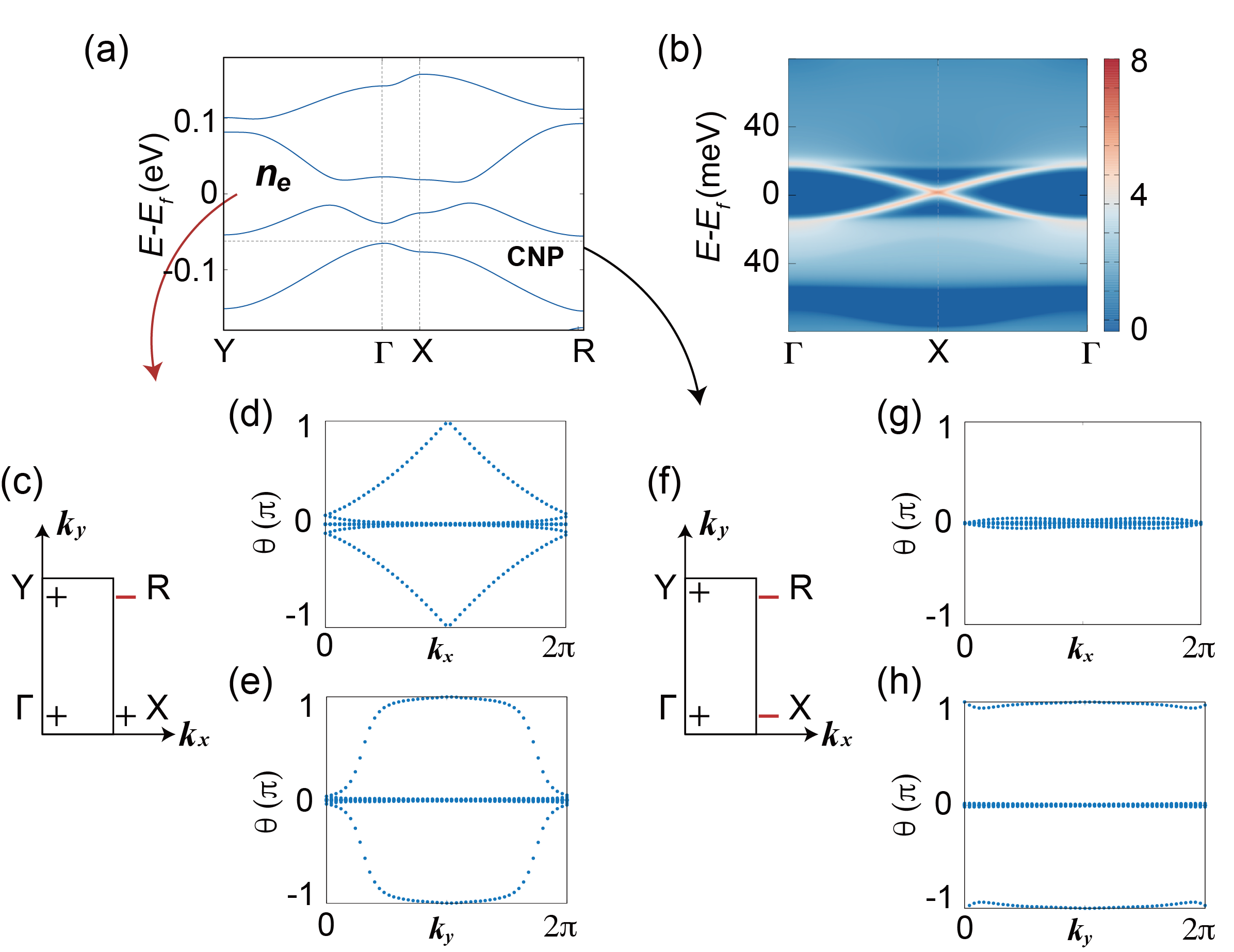}
    \caption{Topological properties for the QSHI + HOTI phase under 2\% tensile strain.
    (a) Band structure showing two distinct gaps at the $n_e$ level and CNP.
    (b) Edge spectral function along the [100] edge, showing helical edge states only at the $n_e$ gap.
    (c) Parity products at four TRIMs for occupied bands up to the $n_e$ gap, giving \(\nu_0 = 1\).
    (d)-(e) Wilson loop of occupied bands up to the $n_e$ gap along \(k_x\) and \(k_y\), respectively.
    (f) Parity products for bands up to the CNP, giving \(\nu_0 = 0\), \(\kappa_1 = 2\).
    (g)-(h) Wilson loop evolution for CNP gap, showing trivial \(\mathbb{Z}_2\) but nontrivial \(\mathbb{Z}_4\) topology.
    }
    \label{fig:s12}
\end{figure}

To further illustrate the impact of strain on correlated topological phases, we present a representative example of the \textit{QSHI + HOTI} phase at 2\% tensile strain, with fixed interaction strength \( U/V_1 = 2.25 \) and dielectric constant \( \epsilon = 13 \). As shown in Fig.~\ref{fig:s12}(a), the band structure reveals two distinct insulating gaps: one at the $n_e$ level and the other at the charge neutrality point. The system thus enters a hybrid phase, where the $n_e$ gap hosts a quantum spin Hall insulator, while the CNP gap transitions into a higher-order topological insulator.

This distinction is confirmed by detailed topological analyses. For the $n_e$ gap, the parity products at the four TRIMs (\(\Gamma, X, Y, R\)) follow the pattern $(+++-)$, as shown in Fig.~\ref{fig:s12}(c), corresponding to a \(\mathbb{Z}_2\) index \(\nu_0 = 1\). The Wilson loop evolution further supports this assignment: along \(k_x\), we observe nontrivial winding with a pair of phases crossing at \(\pi\) (Fig.~\ref{fig:s12}(d)), and a gapless flow of Wannier centers along \(k_y\) (Fig.~\ref{fig:s12}(e)). These features are characteristic of the QSHI phase. Moreover, the helical edge states remain clearly visible within the $n_e$ gap, as shown in Fig.~\ref{fig:s12}(b).

By contrast, the CNP gap exhibits a different topological structure. The parity products follow the distribution \((+-+-)\), indicating that the system undergoes band inversions at both the X and R points. This leads to a trivial \(\mathbb{Z}_2\) index (\(\nu_0 = 0\)), but a nontrivial \(\mathbb{Z}_4\) index \(\kappa_1 = 2\), confirming the HOTI character. The Wilson loops along both directions (Figs.~\ref{fig:s12}(g) and (h)) show flat and gapped phases, with a phase shift of \(\pm \pi\) emerging only along \(k_x\), consistent with boundary-localized Wannier centers typical of HOTIs. No helical edge states are observed in the CNP gap, as expected for a second-order topological insulator.

This result demonstrates that under the same interaction conditions (\( U/V_1 = 2.25 \), \(\epsilon = 13 \)), varying the strain alone can induce a topological phase transition from the previously observed dual QSHI state into a mixed QSHI + HOTI configuration. It also highlights that the CNP gap is more sensitive to strain modulation than the $n_e$ gap, which retains its QSHI topology under small tensile strains.

\begin{figure}[htb!]
    \centering
    \includegraphics[width=0.9\linewidth]{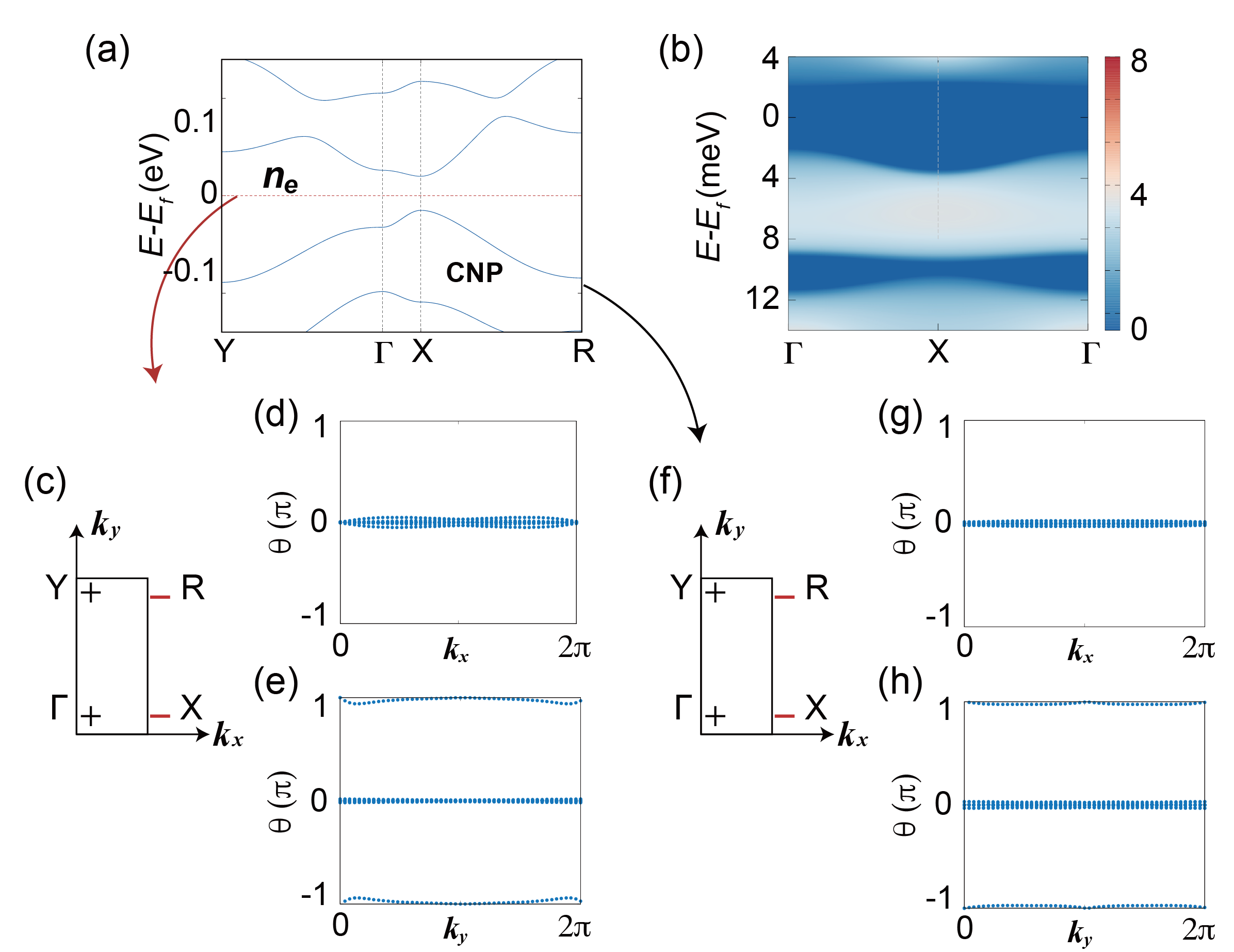}
    \caption{ Topological characterization of the dual HOTI phase under 4\% tensile strain. (a) Band structure for the correlated insulating phases at the $n_e$ and CNP gaps under tensile strain. (b) Calculated edge spectral function along the [100] direction, confirming the absence of helical edge states. (c) Parity products at four TRIM points ($\Gamma$, X, Y, R) for the occupied bands up to the $n_e$ gap, giving a trivial Z$_2$ invariant $\nu_0=0$ but a nontrivial Z$_4$ invariant $\kappa_1=2$. (d-e) Wilson loop evolutions along $k_x$ and $k_y$ directions for the $n_e$ gap, showing nontrivial topology only along the $k_y$ direction. (f) Parity products at TRIM points for occupied bands up to the CNP gap, also giving $\nu_0=0$ and nontrivial $\kappa_1=2$. (g-h) Wilson loop evolutions along $k_x$ and $k_y$ directions at the CNP gap, demonstrating a similar topological behavior as in the $n_e$ gap, thus confirming the presence of a dual HOTI phase.
    }
    \label{fig:dual_HOTI_strain}
\end{figure}

\subsection{Dual HOTI phase under 4\% tensile strain}\label{Ssec:strain_effect_hfmf_2}

By further increasing the tensile strain to 4\%, while keeping the interaction parameters fixed at $U/V_1 = 2.25$ and dielectric constant $\epsilon = 13$, we observe a clear transition from the QSHI + HOTI coexistence at 2\% strain into a robust dual HOTI phase, as summarized in Fig.~\ref{fig:dual_HOTI_strain}. 

In this dual HOTI state, both the $n_e$ gap and the CNP gap exhibit trivial Z$_2$ invariants ($\nu_0=0$), yet share identical, nontrivial Z$_4$ topology characterized by $\kappa_1=2$. Specifically, the parity products at the four TRIM points (\(\Gamma\), X, Y, R) indicate band inversions occurring simultaneously at both the X and R points for both $n_e$ and CNP gaps, reflecting double band inversions that stabilize higher-order topological insulator phases.

Further confirmation is provided by Wilson loop analyses. As shown in Fig.~\ref{fig:dual_HOTI_strain} (d-e,g-h), Wilson loop evolutions along $k_x$ exhibit trivial phases without crossings, whereas along $k_y$ direction, nontrivial $\pm\pi$ phase shifts occur consistently in both gaps. Such anisotropic Wannier center evolutions precisely match the parity products analysis, verifying the Z$_4$ topology. Moreover, the edge spectral function along the [100] edge (Fig.~\ref{fig:dual_HOTI_strain}(b)) clearly shows the absence of gapless helical edge states, consistent with the HOTI classification. 

Thus, the 4\% tensile strain scenario provides clear evidence that mechanical strain acts as an effective tuning parameter, driving the system from QSHI or mixed phases into stable, purely HOTI phases, further enriching the topological phase diagram of correlated TaIrTe$_4$.

\section{Experimental Details} \label{Sec:exp}
\subsection{TaIrTe$_4$ Crystal Growth Process:} \label{Ssec:Exp-growth}

TaIrTe$_4$ crystals were synthesized using the Te-flux method with a molar ratio of Ta:Ir:Te ( $>$ 99.99$\%$ purity) of 3:3:94. The elements were placed in an alumina crucible, sealed in a vacuum quartz tube, and gradually heated to 1373 K, where they were held for 10 hours. The mixture was then slowly cooled to 873 K at a rate of 2 K/h. The resulting TaIrTe$_4$ crystals were separated from the excess Te flux via centrifugation.

\begin{figure*}[ht] 
    \centering
    \includegraphics[width=0.7\textwidth]{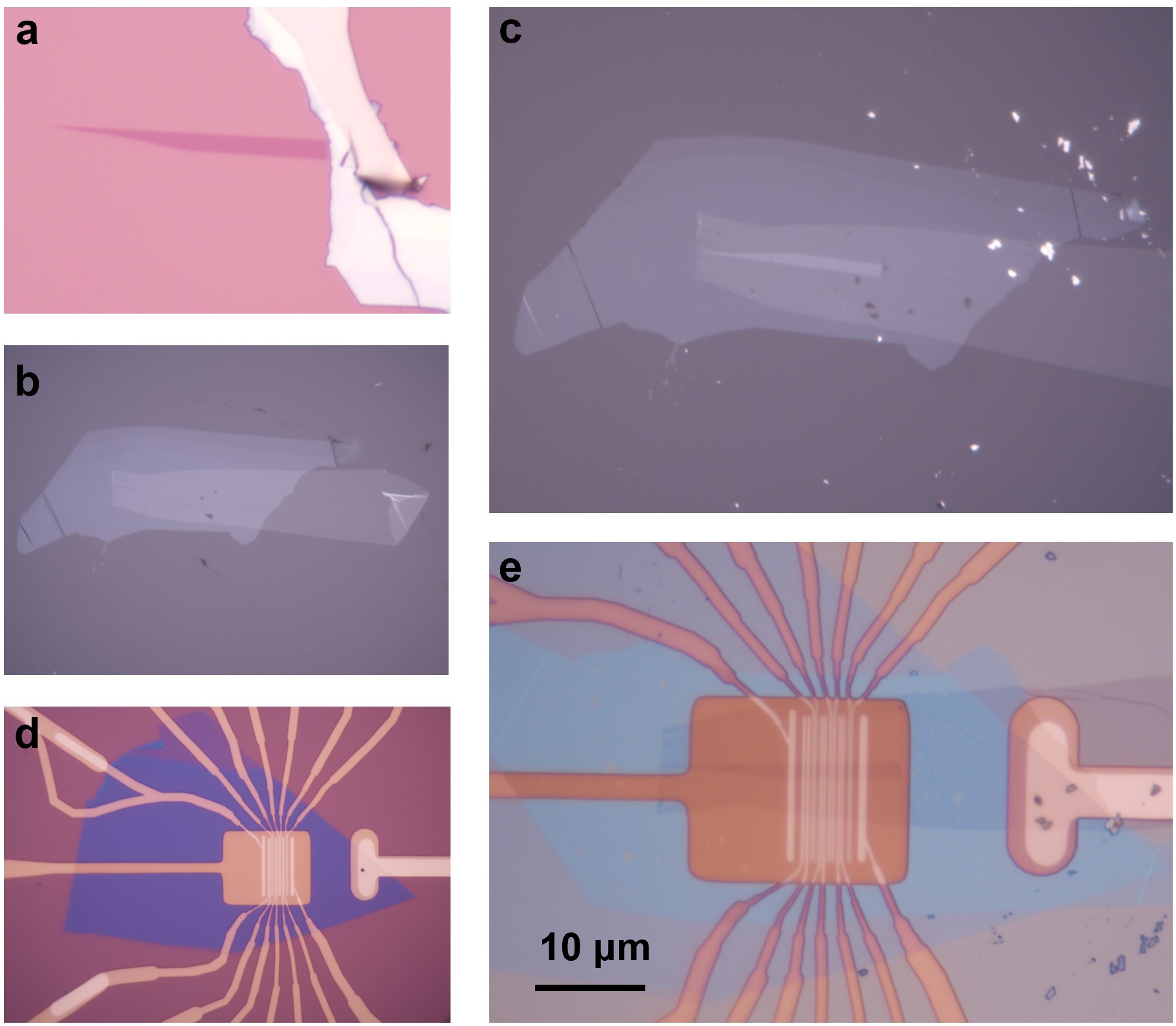}
    \caption{Device fabrication process of monolayer TaIrTe$_4$ device.
    (a) Optical image of an exfoliated monolayer TaIrTe$_4$ flake with a ribbon-like shape on a SiO$_2$ substrate. The exfoliation was performed in a glovebox under an inert atmosphere.
    (b) A stack consisting of few-layer graphene (FLG, serving as the top gate) and boron nitride (BN) is prepared on a PC/PDMS stamp. 
    (c) The monolayer TaIrTe$_4$ flake is picked up using the FLG/BN stack.
    (d) Optical image of the bottom-contact structure, where a metal pad serves as the back gate, and BN acts as the dielectric layer. Contact electrodes are patterned on the surface, with a small gap between adjacent electrodes to define the channel length.
    (e) Optical image of the final device structures.
    }
    \label{fig:fab}
\end{figure*}

\subsection{TaIrTe$_4$ Exfoliation and Device Fabrication:} \label{Ssec:Exp-device}
TaIrTe$_4$ flakes were exfoliated onto oxygen plasma-cleaned SiO$_2$ chips using the scotch-tape method inside an Ar-filled glovebox to solve the air sensitivity nature of thin flakes. The layer thickness was determined through optical microscopy (Fig.~\ref{fig:fab}(a) shows a monolayer TaIrTe$_4$ flake). Target flake regions were patterned using a tip-scratching and cleaning process. For subsequent short-channel device fabrication, stripe-shaped TaIrTe$_4$ flakes with naturally exfoliated boundaries were preferentially selected.\\

We designed a bottom-contact, dual-gated device structure to encapsulate the monolayer TaIrTe$_4$ flake and enhance gate tunability. First, few-layer graphene (FLG) and boron nitride (BN) flakes were sequentially picked up using a PDMS/PC stamp to serve as the top gate and dielectric layer, respectively (Fig.~\ref{fig:fab}(b)). Next, the monolayer TaIrTe$_4$ flake was picked up by the stack (Fig.~\ref{fig:fab}(c)). Subsequently, bottom-contact electrodes (Ti 2 nm/Pt 10 nm) were patterned onto the BN and metal back-gate surface (Fig.~\ref{fig:fab}(d)), with the channel length defined by the spacing between adjacent electrodes. The bottom-contact structure was then thermally annealed in a forming gas atmosphere (Ar:H$_2$ = 1:1) at 350 $^\circ$C for 3 hours to remove polymer residues and surface contaminants. Finally, the FLG/BN/TaIrTe$_4$ stack was precisely aligned and transferred onto the bottom-contact structure using a transfer stage, ensuring that the FLG was in contact with the pre-patterned top-gate electrode. The final device structure is shown in Fig.~\ref{fig:fab}(e).\\

\subsection{Electrical measurements of TaIrTe$_4$ devices:}

Electrical measurements were performed using Montana and Cryomagnetics cryostats. The electrical transport properties of TaIrTe$_4$ devices were characterized using standard lock-in techniques, and gate voltages were applied via Keithley source meters. In dual-gated devices, the carrier density ($n$) was controlled through a combination of the top-gate voltage ($V_\textrm{tg}$) and bottom-gate voltage ($V_\textrm{bg}$).

\begin{equation}
n=\frac{\epsilon_{\textrm{0}}\epsilon_{\textrm{BN}}V_{\textrm{bg}}}{ed_{\textrm{b}}}+\frac{\epsilon_{\textrm{0}}\epsilon_{\textrm{BN}}V_{\textrm{tg}}}{ed_{\textrm{t}}},
\label{n_equation}
\end{equation}

Here, $\epsilon_{\textrm{0}}$ is the vacuum permittivity, $\epsilon_{\textrm{BN}}$ is the BN dielectric constant ($\epsilon_{\textrm{BN}}\approx 3$) and $d_\textrm{b}(d_\textrm{t})$ is the dielectric thickness of the bottom (top) BN.

\subsection{Thermal Activation Gap Analysis}\label{Ssec:thermal_gap}

\begin{figure*}[htb!] 
    \centering
    \includegraphics[width=0.8\textwidth]{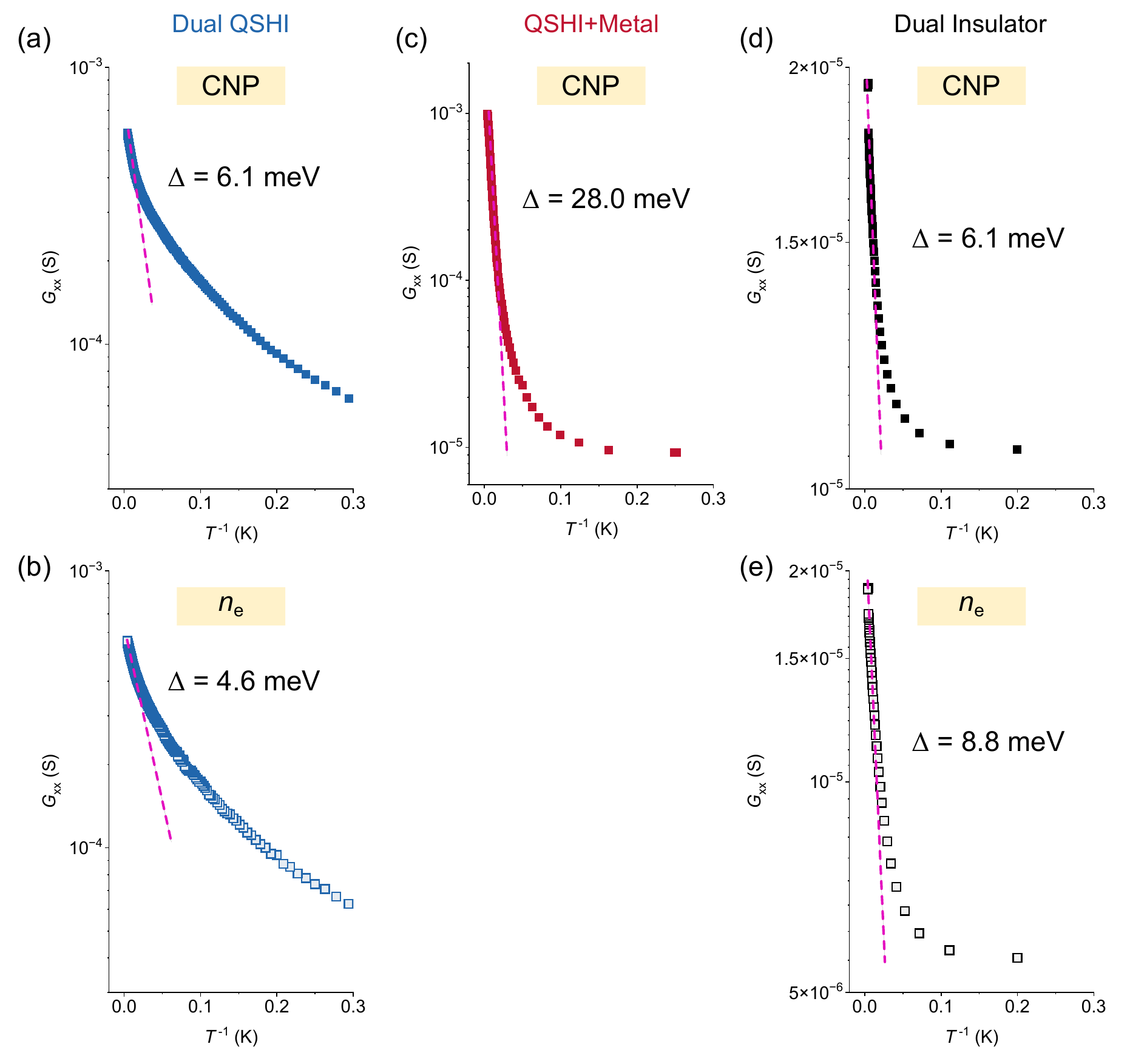}
    \caption{Band gap extraction via thermal activation fitting. The activation gaps are extracted based on the temperature-dependent conductance $G_{xx}$ by using the Arrhenius relation $G_{xx} \propto \exp(-\Delta/2k_B T)$. (1) dual QSHI: (a) $\Delta_{CNP}=6.1$ meV; (b) $\Delta_{n_e}=4.6$ meV; 
    (2) QSHI+Metal: (c) $\Delta_{CNP}=28.0$ meV; 
    (3) dual Insulator: (d) $\Delta_{CNP}=6.1$ meV; (e) $\Delta_{n_e}=8.8$ meV; 
    }
    \label{fig:thermal_gap}
\end{figure*}

To quantitatively evaluate the insulating behavior observed at the charge neutrality point and on the electron-doped side, we perform thermal activation analysis on the longitudinal conductance $G_{xx}(T)$ for three representative monolayer TaIrTe$_4$ devices. As shown in Fig.~\ref{fig:thermal_gap}, the temperature dependence of $G_{xx}$ (extracted from the data shown in Fig.~\ref{fig:typical transport behaviors}) is fitted using the Arrhenius relation:
\begin{equation}
G_{xx}(T) \propto \exp(-\Delta/2k_B T),
\end{equation}
where $\Delta$ denotes the activation energy gap, $T$ is the temperature, and $k_B$ is the Boltzmann constant.
Upper panel of Fig.~\ref{fig:thermal_gap} shows $G_{xx}$ at the CNP for three distinct device types. The extracted thermal gaps are:
$\Delta_{CNP} = 6.1$ meV (dual-QSHI);
$\Delta_{CNP} = 28.0$ meV (metal + QSHI);
$\Delta_{CNP} = 6.1$ meV (dual-insulator).
Lower panel shows the thermal activation behavior on the electron-doped side 
at $n =$ $n_e$:
$\Delta_{n_e} = 4.6$ meV (dual-QSHI);
$\Delta_{n_e} = 8.8$ meV (dual-insulator).

These results support our classification of correlated phases based on edge transport, and confirm that dual-QSHI and metal+QSHI phases are associated with larger, robust gaps, while the dual-insulator phase shows a much weaker insulating behavior consistent with non-topological or higher-order insulating states.








\subsection{Statistical significance and experimental reproducibility}

\begin{figure}[ht]
\centering
\includegraphics[width=0.8\textwidth]{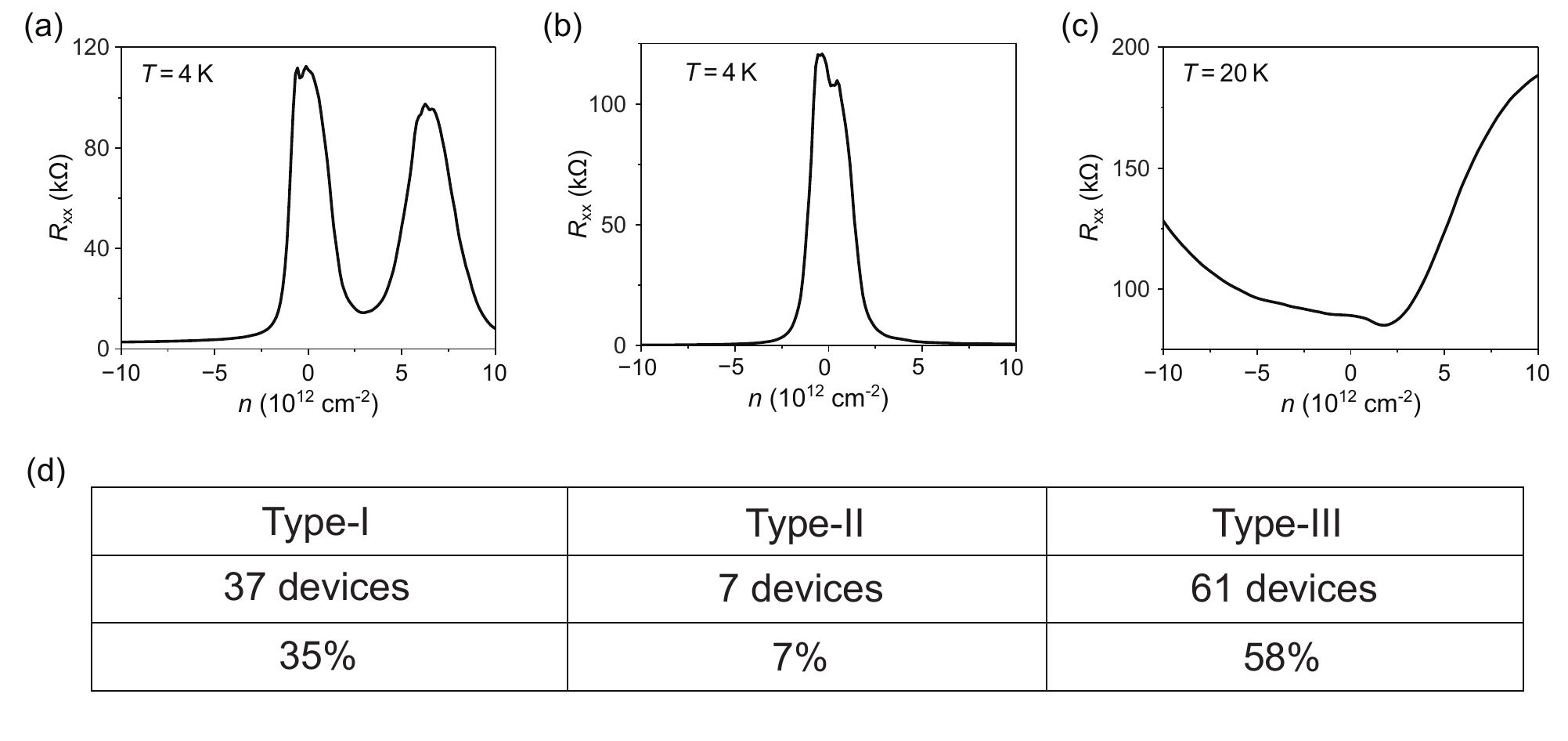}
\caption{Statistical classification of monolayer TaIrTe$_4$ devices based on transport signatures. Out of 105 high-quality devices, three distinct types are identified based on the resistance behavior $R_{xx}(n, T)$ as a function of carrier density $n$ and temperature $T$. Top panels: Representative $R_{xx}$–$n$ curves at low temperature ($T=4$–$40$ K) for each device type. 
Bottom: Summary table listing the number and percentage of devices in each category:  Type-I (\textit{dual QSHI}), Type-II (\textit{QSHI + metal}), and Type-III (\textit{dual insulator}).
}
\label{fig:device_stats}
\end{figure}

To systematically assess the reproducibility and universality of the observed correlated topological phases, we performed transport measurements across a large number of monolayer TaIrTe$_4$ devices. A total of 105 high-quality devices were investigated, allowing robust statistical analysis of the observed transport behaviors.

Based on the characteristic $R_{xx}(n,T)$ profiles, we identified three representative device categories:

\begin{enumerate}
    \item \textbf{Type-I (\textit{dual QSHI})}: Comprising about 35\% of measured devices, these samples exhibit quantized edge conductance at both the CNP and an electron-doped correlated insulating state, consistent with the theoretically predicted dual QSHI phase arising from significant electron-electron interactions near the vHS.

    \item \textbf{Type-II (\textit{QSHI + Metal})}: Representing approximately 7\% of the devices, these samples exhibit well-quantized edge conductance only at the CNP and metallic behavior at finite doping, indicative of relatively weak electron correlations insufficient to stabilize correlated insulating phases at the van Hove singularities.

    \item \textbf{Type-III (\textit{dual insulator})}: The majority (58\%) of measured devices demonstrate insulating transport throughout the doping range, without clear edge conductance quantization. This category likely encompasses trivial insulating or higher-order topological insulating phases, both characterized by the absence of gapless helical edge modes. The prevalence of this category suggests that minor variations in sample quality, device disorder, gating uniformity, or local strain conditions significantly affect the electronic structure, potentially suppressing or obscuring clear quantization.
\end{enumerate}

The experimental distribution highlights two key insights. First, the considerable proportion (35\%) of Type-I devices confirms that the interaction-induced dual QSHI phase is robustly reproducible and experimentally accessible. However, the dominance of Type-III devices underscores practical challenges in reliably achieving ideal topological phases, indicating a strong sensitivity of the correlated states to disorder, gating uniformity, and subtle variations in fabrication process.

Overall, the comprehensive device statistics validate the experimental realization of theoretically predicted correlated topological phases and emphasize the critical importance of precise control over material quality and device conditions for future investigations and applications of correlated topological materials.




\end{document}